\documentclass[twocolumn]{aastex63}
\usepackage{epsfig,graphicx}
\usepackage{url}
\usepackage{hyperref}

\received{2019 October 24}
\revised{2019 December 21}
\accepted{2020 January 3}

\acceptjournal{ApJ}
\setwatermarkfontsize{0.7in}

\shorttitle{Testing the solar activity paradigm in the context of exoplanet transits}
\shortauthors{Schrijver}

\begin{document}
\title{Testing the solar activity paradigm in the context of exoplanet transits}
\author[0000-0002-6010-8182]{Carolus J.\ Schrijver}
\affiliation{Johannes Geiss Fellow at the International Space Science
  Institute, \\Hallerstrasse 6, 3012 Bern, Switzerland}
\email{karelschrijver@gmail.com}

\correspondingauthor{C.J.\ Schrijver}

\nocollaboration{2}

\begin{abstract}
  Transits of exoplanets across cool stars contain blended information
  about structures on the stellar surface and about the planetary body
  and atmosphere. To advance understanding of how this information is
  entangled, a surface-flux transport code, based on observed
  properties of the Sun's magnetic field, is used to simulate the
  appearance of hypothetical stellar photospheres from the visible
  near 4000\,\AA\ to the near-IR at 1.6\,$\mu$m, by mapping
  intensities characteristic of faculae and spots onto stellar
  disks. Stellar appearances are
  computed for a Sun-like star of solar activity up to a star with
  mean magnetic flux density $\sim 30\times$ higher. Simulated transit
  signals for a Jupiter-class planet are compared with
  observations. This
  (1) indicates that the solar paradigm is
  consistent with transit observations for stars throughout the activity range
  explored, provided that infrequent large active regions with fluxes
  up to $\sim 3\times 10^{23}$\,Mx are included in the emergence spectrum,
  (2) quantitatively confirms that for such a model, faculae brighten
  relatively inactive stars while starspots dim more-active stars, and suggests
  (3) that large starspots inferred from transits of active stars
  are consistent with clusters of more compact spots seen in the model
  runs,
  (4) that wavelength-dependent transit-depth effects caused by
  stellar magnetic activity for the range of activity and the
  planetary diameter studied here can introduce apparent changes in
  the inferred exoplanetary radii across wavelengths from a few
  hundred to a few thousand kilometers, increasing with activity, and
  (5) that activity-modulated distortions of broadband stellar
  radiance across the visible to near-IR spectrum can reach several
  percent.
\end{abstract}


\section{Introduction}\label{sec:intro}
Magnetic activity of cool stars, as in the case of the Sun, manifests
itself in a photosphere in the form of a mixture of bright faculae and
dark pores and spots. Their distribution and number is modulated by
the stellar dynamo, which in the case of the Sun is reflected in the sunspot
cycle. Knowledge of the \replaced{geometric properties}{locations and
  sizes} of active regions and their spots on stars other than the Sun
is limited by the means available to study them, generally through the
modeling of photometric modulation \replaced{associated with stellar
  rotation or the analysis of (Zeeman) Doppler signals. These methods
  are hampered by the intrinsic evolution of the activity structures
  as the star rotates, by line-broadening effects due to thermal and
  convective motions, and generally by a north-south hemispheric
  ambiguity. The result is that often significant simplifications are
  made and regularization algorithms applied}{or of (Zeeman) Doppler
  signals associated with stellar rotation. These methods generally
  have to assume that the structures on the stellar surfaces do not
  evolve on the time scale of a rotation; that clearly is not the case
  for individual active regions on a star with a rotation rate like
  that of the Sun, nor is it likely to hold on faster rotators where
  the rate of flux emergence is significantly elevated. Doppler
  signals in principle carry information on the latitude of the source
  regions, but this is only recoverable if the Doppler signal of
  rotation is comparable to, or exceeds, that of the thermal and
  convective velocities; for a star like the Sun, with a photospheric
  thermal line width equivalent to some 7\,km/s, the maximum
  rotational Doppler signal exceeds the thermal width only for stars
  with rotation periods below $\approx 7$\,days. For stars with a
  rotation axis near the plane of the sky, both hemispheres are
  accessible to these methods, but none can disambiguate on which
  hemisphere the features reside that lead to the observed
  signals. Magnetic field maps are even harder to reconstruct than
  spot maps; for a discussion of methods and results, see the review
  by \citet{2012LRSP....9....1R}.

Because of these intrinsic challenges, only large-scale,
slowly evolving structures can in principle be recovered with some
reliability, as significant
simplifications are made by the models and regularization algorithms applied} to
limit degeneracies of possible but not necessarily plausible
solutions. Thus, angular resolution is limited to large patches
commonly spanning tens of astrocentric degrees; for example,
\citet{2007A&A...464..741L} find that the longitudinal resolution
achievable through spot modeling of rotational modulation of a star of
solar activity is typically some $60^\circ$, while
\citet{2009AnARv..17..251S} notes that the smallest ``spots'' found
under optimal conditions for Doppler imaging are some
$10^\circ\times 10^\circ$, the latter just reaching the size of the
largest (rare) sunspots. \added{For slow rotators, with
  $v\,\sin(i)\le 5$\,km/s which are the focus of the present study,
  however, the resolution achievable by Zeeman-Doppler imaging is
  limited to some $30^\circ$ while much of the flux in a map is lost
  as signals with opposite polarizations largely cancel \citep{2019MNRAS.483.5246L}.}

Despite the limitations of how stellar magnetic activity can be
studied, much has been learned about activity patterns (such as the
apparent existence of persistently active longitudes) and stellar
differential rotation (albeit commonly with ambiguity of sign). But
the emergence patterns as a function of time and latitude, the size
distribution of active regions, the magnitude of convective flux
dispersal, the strength and profile of meridional advection, and
properties of the full spectrum of starspots remain largely beyond our
observational grasp.  For reviews on stellar dynamos, starspots, and
stellar magnetic fields, see, for example,
\citet{2005LRSP....2....8B}, \citet{2017LRSP...14....4B},
\citet{2012LRSP....9....1R}, and \citet{2009AnARv..17..251S}.

The discovery of transiting exoplanets has opened up a novel way of
studying stellar surface structures: the transiting exoplanet acts as
an occulting disk moving across the features within the transit
path. The spatial resolution that can be attained by this method is in
principle limited only by the size of the occulting disk and its
movement across the stellar disk within the time required for an
exposure with sufficient signal-to-noise ration (S/N). If the transit is
sufficiently fast, the background photospheric features can be assumed
constant, so that difference spectra (or passband intensity
differences) can be computed between the unocculted and occulted
signals to acquire a pure spectrum of what resides behind the
exoplanetary disk \citep[see][for an example of studying stellar
granulation]{2017A&A...605A..90D}. For slower transits, or to obtain
higher resolution, the time series of the transit can be studied as a
convolution problem. Light-curve analyses suggest spot sizes down to
$\sim 3^\circ$ in diameter \cite[{\em
  e.g.},][]{2011AnA...529A..36S,2017ApJ...835..294V}.

This fairly novel way to achieve high-resolution information for
stellar surface structures along the transit path is of critical
importance to studying exoplanetary atmospheres through transit
spectroscopy \citep[{\em e.g.},][]{2000ApJ...537..916S,2001ApJ...553.1006B}:
the light passing through the exoplanetary atmosphere contains
information about the atmospheric structure, but to extract it
unambiguously, the properties of the light source need to be
known. That light source is the photospheric patch that lies behind
the exoplanetary atmosphere in the line of sight towards the observer,
as discussed in detail by, {\em e.g.}, 
\citet{2018ApJ...853..122R,2019AJ....157...96R}, although some scattering
in the atmosphere of the exoplanet needs to be allowed for. This
presents a challenge and opportunity combined: the emission properties
of the stellar surface structures along the transit trajectory as well
as the transmission properties of the exoplanetary atmosphere need to
be analyzed in a joint study of the spectrum that carries information
on both source and absorber
\citep[{{\em e.g.,}}][]{2019BAAS...51c.149K,2019BAAS...51c.328R}.

The stellar magnetic activity can affect observed spectra whose
interpretation can lead to ambiguities between stellar surface (and,
of course, chromospheric) features and the chemical composition, clouds,
and particulate hazes in an exoplanetary atmosphere, arising in the
interpretation of broadband measurements or high-resolution line
spectroscopy \citep[see, for example,][]{stellaractivitytransitspectra,AURAtransmissionspectroscopy}.

Until recently, transit spectroscopy studies often assumed that the
transit light source is the same as the spectrum of the average
unocculted disk. More recently, detailed shapes of limb darkening
\citep[{\em e.g.},][]{2015MNRAS.450.1879E} and modeling of spots along the
transit path have come into play, with an occasional study also
allowing for bright facular regions \citep[{\em e.g.},][and others, see
below]{2014AnA...568A..99O,2019A&A...622A.172M,2019arXiv191105179B},
\added{albeit characerized by a simple brightness contrast or single
  temperature without accounting for limbbrightening effects}.
Transit light-curve analyses in the literature go about inferring
properties of stellar surface structures from observations, generally
assuming simplified features on the stellar surface: often simply
circular starspots, with or without a variety of brightnesses (or
assumed spot temperatures), some with associated facular regions
(which may all be assumed to be identical except for sizes, or may
allow for some simple ad hoc spectral effects), either surrounding
these spots or allowing patches elsewhere on the stellar
surface. Understandably, the number of free parameters is kept low
because present-day observations can offer only a limited number of
constraints.

The purpose of this paper, in contrast, is to use forward modeling of
patterns of stellar magnetic fields and their photospheric effects
based on observed characteristics of the Sun's magnetic field. Here,
too, even if limiting this study to stars like our G2\,V Sun, there is
an assumption, and indeed one that has many facets, namely that the
solar paradigm applies to other stars, and specifically to stars of
substantially higher magnetic activity than exhibited by the Sun. This
study is thus meant as an exploration through attempted falsification
of the validity of the solar example. At the same time, it provides
quantitative insight into the impacts of stellar activity on
information about exoplanetary atmospheres extracted from transit
spectroscopy. Among the multitude of choices for such a study, I
choose to use the full richness of surface field patterns and
model-derived limb-brightening curves for faculae in areas of
different mean magnetic flux density, while stopping short of full
spectral synthesis. For approaches in which the full observed
spectrum is analyzed or a model spectrum
computed but with highly simplified assumptions about stellar surface
features, see, for example (ordered in time, generally following
a path toward increasing complexity) 
\citet{2008MNRAS.385..109P,2011MNRAS.416.1443S,2016AnA...586A.131H,2018ApJ...853..122R,2019AJ....157...96R}.

\begin{table*}
  \caption{\label{tab:transits} Selection of active G- and early-K type
    dwarf stars transited by Jupiter-sized exoplanets. Listed are:
    name, spectral type, rotation period, exoplanetary
    radius $R_{\rm p}$ relative to the stellar radius $R_\ast$, the
    estimated relative impact parameter for the transit $d_{\rm t}$, and
    characteristic \replaced{peak-to-peak}{peak-to-trough} amplitude of the rotational
    modulation of the stellar brightness. The next-to-last column
    shows whether evidence is seen in the transit residuals for
    surface inhomogeneities (with estimated maximum magnitude
    if seen, or the detection threshold if not). }
  \begin{tabular}{llrlrrll}
    \hline 
Star & Spectral & $P_{\rm rot}$ & $R_{\rm p}/R_\ast$ & $d_{\rm t}\,\,^d$ &Rotation & Transit &   Reference(s) \\
& type       & (days)           & &
                                                         & modul.
                                                                    & features? & \\
    \hline 
V1298\,Tau & K0--K1.5 & $2.87$& $0.071$ &0.24&$\approx 3$\%\ & Y$(\sim 0.15\%)$ &\citet{2019AJ....158...79D} \\
CoRoT-2 & G7\,V & $4.5$ & $0.172$ &0.25&$\approx 3-4$\%\ & Y$(\sim 0.6\%)$& \citet{2011AnA...529A..36S} \\
POTS-1 & K5\,V& $\ga 7$ $^a$& $0.164$&0.45 &?& N$(\la 0.7\%)$ &\citet{2013MNRAS.435.3133K}\\
HAT-P-23 & $\approx$G1 & $\approx 7$ $^b$& $0.13$&0.33&?& N$(\la 0.2\%)$ & \citet{2011ApJ...742..116B}\\
WASP-170 & G1\,V& $7.8$ & 0.118 &0.67&$\approx 2$\%\ & Y$(\sim 0.3\%)$& \citet{2019AJ....157...43B}\\
CoRoT-9 & G3\,V& 8 $^b$& 0.115&0.16 &?& Y$(\sim 0.2\%)$ & \citet{2017AnA...603A.115L}\\
    WASP-36 & G2\,V& $\approx 11$$^c$& 0.138&0.66&?& N$(\la 0.05\%)$& \citet{2016MNRAS.459.1393M}\\
Kepler-17 & G2\,V& 12 & 0.138 &0.26&$\approx 1-3$\%\ & Y$(\sim 0.3\%)$&
                                                                   \citet{2017ApJ...835..294V},  \\
     &&&&&&&\citet{2019AnA...626A..38L}\\
   
HD\,189733 & K1--K2\,V& 12& 0.144&0.65 &$\approx 1.5$\% & Y$(\sim 0.3\%)$&
                                                          \citet{2011MNRAS.416.1443S} \\
WASP-52 & K2\,V& 16--18& 0.167&0.74&$\approx 1$\%& Y$(\sim 0.3\%)$& \citet{2013AnA...549A.134H},\\
     &&&&&&& \citet{2019arXiv191105179B}\\
Kepler-71 & G7--G9\,V & 20 & 0.136 &--& $\approx 1$\%\ & Y$(\sim 0.3\%)$ & 
                                                                        \citet{2019MNRAS.484..618Z} \\
WASP-6 & G8\,V & 24 & 0.141 &0.28& $\approx 1$\%\ &  Y$(\sim 0.3\%)$ &
                                                                  \citet{2015MNRAS.450.1760T},\\
    &&&&&&&\citet{2015MNRAS.447..463N} \\
    \hline \\
  \end{tabular}
  
  $^a$ Based on $v \sin i \le 5.3$\,km/s, and an estimated radius of
  $0.7 R_\odot$; $^b$ based on $v \sin i=5.4$\,km/s in
  the SIMBAD database; $^c$ based on $v \sin i=4$\,km/s in
  the SIMBAD database; $^d$ estimated from orbital inclination,
  orbital semi-major axis, and stellar radius as listed in the tables
  at exoplanet.eu.
\end{table*}
The hypothesis that the solar paradigm and the derived surface
flux-transport model hold for stars of different activities has been
extensively tested \citep[{\em e.g.,}][]{schrijver+zwaan99}. The model
applied here was used to understand how solar and cool-star radiative
losses from chromospheres and coronae are related
\citep{schrijver2000}, supporting the applicability of the solar
paradigm for stellar studies. For the most active stars, the phenomena
of extremely large and high-latitude starspots present challenges to
the solar paradigm \citep[but see][]{schryver+title2001}, and for this
reason the present study stops at stars of moderate activity. Among
the issues that will come up is the phenomenon of the starspot: light
curves of active stars often suggest starspots much larger than
sunspots. However, the low resolution that can be
achieved when mapping stellar surface features leaves open the
possibility that such large stellar spots are in fact clusters of
smaller, more solar-like ones, which at least for moderately active
stars appears to be compatible with stellar data
\citep[{\em e.g.},][]{2004MNRAS.348..307S}, and also with lifetime
estimates of starspots assuming these are dispersed subject to surface
convective flows \citep{2014ApJ...795...79B}.

In this paper, I experiment with a flux-transport model developed for
the Sun, discussed in Sect.~\ref{sec:transport}. Model stellar
radiances are computed using a three-component approach that includes
a quiet background photosphere, a component associated with magnetic
field in magnetic plages and network (referred to here collectively as
faculae), and spots (including their smaller counterparts, the pores).
Modeling of the solar irradiance has shown such an approach to be
successful, see, {\em e.g.},
\citet{1998A&A...335..709F,2017PhRvL.119i1102Y}. Properties of solar
faculae, pores, and spots, and a quiet-Sun limb-darkening curve are
then used to render the appearance of Sun-like stars of different
activity at two different wavelengths in the visible and one in the
near-IR.  Subsequently, virtual exoplanet transits are performed to
compute the color-dependent transit curves and residuals
(Sect.~\ref{sec:observations}). Simulated rotational modulation
signals (Sect.~\ref{sec:rotmod}) and transit residuals
(Sect.~\ref{sec:transits}) are then discussed and compared to examples
of observed signals in a few cool dwarfs (presented in
Sect.~\ref{sec:sample}), before I conclude the paper with discussion and conclusions
(Sects.~\ref{sec:discussion} and~\ref{sec:conclusions}).

\section{A selection of cool stars transited by large
  exoplanets}\label{sec:sample}
The model discussed in Sect.~\ref{sec:model} relies on observed
properties of the Sun's surface activity, including the shape of the
sunspot cycle, the frequency distribution of active-region sizes,
their distribution to form the butterfly diagram, their typical tilt
angles relative to the equator, their nesting property for successive
generations of emergence, the (super-)granular random-walk dispersal,
and the large-scale meridional advection and differential
displacement. None of these properties are known well enough for other
types of stars (be it from observations or through modeling), so that
application of the model to anything but an early G-type dwarf star is
not warranted without determination of changes in the above properties
with spectral type. Similarly, for the generation
of transit light curves, note that the wavelength and
position-dependent properties of sunspots and faculae are known for
the Sun, but are insufficiently clear for other types of stars. In view of all
that, the model is applied here only for a Sun-like star, and
consequently the comparison to observations is limited to G- and early
K-type main-sequence stars.

Table~\ref{tab:transits} lists a selection of relatively active G- and
early K-type cool dwarf or subgiant stars that are transited by giant
exoplanets for which transit light curves and transit residuals have
been analyzed in the literature. The table lists a few of the stellar
properties, followed by characteristic
\replaced{peak-to-peak}{peak-to-trough} values of the relative
rotational modulation; \added{the latter are rough estimates based on
  observed light curves, averaging over a series of about a dozen
  rotations}. There are also notes on whether there are significant
signatures of stellar surface features seen in the transit residuals,
and their peak relative strength.  A few comments on three of these
(with source references as in the table, unless otherwise given):
\begin{itemize}

\item CoRoT-2 is a rather young and rapidly spinning star. Photometry
  of 77 transits over $\approx 30$ full stellar rotations
  suggested typically five dark patches interpreted as spots per
  transit. The patch sizes measure typically
  $0.25$--$0.8$ planetary radii ($R_{\rm p}$), averaging at
  $\approx 0.45 R_{\rm p}$, $0.077 R_\ast$, or $\approx
  50,000$\,km. The relative spot intensities exhibit a wide range, almost
  uniformly covering values from close to zero up to around 0.8,
  falling rapidly in frequency above that. Figure~4 by
  \citet{2011AnA...529A..36S} suggests a pronounced tendency for the
  smaller spots to be darker. As discussed in
  Sect.~\ref{sec:transits}, this could indicate that the larger spots
  are in reality clusters of dark structures with an area filling
  factor decreasing with increasing feature size (this relationship is
  not explicitly shown, but the figures suggest the scaling may be
  roughly linear). Rotational modulation reaches peak-to-trough values of
  $3$\%--$4$\%, \citep{2009A&A...493..193L}.

\item Kepler-17 is a Sun-like star, albeit more active at $P_{\rm
  rot}\approx 12$\,days. Transit light-curve analyses over a period of
some 1200\,days, with residuals
reaching $\approx 0.4$\%, resulted in the determination of spot properties under the
transit chord. Typical spot radii range from 0.2 to 0.7$R_{\rm p}$,
with a few larger ones; intensities range from about 0.2 to 0.85, with
an average value of 0.55. As in CoRoT-2, there is a trend for larger
spots to be brighter.


\item Kepler-71 is a late-G main-sequence star with
  $P_{\rm rot}=20$\,days. Its transit residuals have been modeled in
  terms of both dark spots and bright faculae. The modeling suggests
  spot intensities widely ranging between about 0.1 and 0.9 times the
  unperturbed photosphere, while facular areas have relative
  intensities most commonly between 1.1 and 1.25 in the {\em Kepler}
  passband. The spots are typically found to have a scale of half the
  exoplanet radius, while the facular regions range from about that
  size up
  to about 1.3$R_{\rm p}$, covering roughly 50--100\%\ more area in
  total than the spots do.
\end{itemize}

\section{Model description}\label{sec:model}
\subsection{Flux-transport model}\label{sec:transport}
The flux-transport model used here was developed by
\citet{schrijver2000}, with cycle modulation introduced by
\citet{schryver+title2001} and with flux-decay properties as discussed
by \citet{schrijver+etal2002a}. This same flux-transport code
continues to be used in an assimilation mode for SOHO/MDI and SDO/HMI
magnetogram data starting in mid-1996, see, for example,
\citet{schrijver+derosa2002b}, and
\citet{schrijver+liu2008}.\footnote{Assimilation results can be viewed
  at http://www.lmsal.com/forecast/.}\added{\footnote{Results from another
    flux-transport model are shown by \citet{2016MNRAS.456.3624G} and
    \citet{2019MNRAS.483.5246L}, in which also experiments with
    different differential-rotation and meridional-advection profiles
    are discussed. The main differences with the model used here are
    that their model describes the magnetic field as a
    continuous medium, does not reach down to the smallest bipoles,
    imposes a flux-independent diffusion coefficient, and has a
    distinct flux insertion algorithm that moreover does not
    incorporate active-region nesting.}}

Visualizations of the global field patterns are discussed below,
but other realizations associated with two of the model runs here
(identified as ${\cal M}({\cal A}=1)$ and ${\cal M}({\cal A}=30)$; see
below in this section for their properties) can be found in
\cite{schrijver+etal2003}. 

At each time step, a set of bipolar regions is randomly selected from
a parent-distribution \replaced{frequency spectrum.}{power-law
  frequency spectrum
  that is based on the work by \citet{1993SoPh..148...85H}: they used
  magnetogram sequences throughout solar
    Cycle~21 (1975--1986) to determine times of
  maximum development of bipolar regions and with that determined the
  behavior of the flux-input spectrum with cycle phase. Their findings
  support a power-law distribution of bipolar-region sizes for which
  the power-law index is essentially unchanged through the cycle,
  so that only a multiplier is needed to describe the evolution
  through the activity cycle.}
The same spectrum is used throughout the simulated activity
cycles, but modulated in time with a profile resembling a solar
sunspot cycle, with successive cycles somewhat 
overlapping in early and late cycle
phases. Stars of different activity levels are simulated by a
multiplier on the cycle amplitude. As discussed below, a maximum flux
for active regions is set as desired, below which the distribution is
maintained the same in shape, except for an overall multiplicative factor.

These regions are then randomly distributed over a range of latitudes
and with a range of tilt angles. \added{The latitudes and tilt angles
  have a spread that decreases with  increasing size of the region, while the mean
  latitude shifts from mid-latitudes toward the equator as a cycle
  progresses, with a 3\,yr phase of overlap between successive
  cycles, all based on the average values and scatter about those
  derived from solar observations.} Random values are generated, drawn
from parent distribution functions that are based on observed
properties of solar bipolar regions, ranging from small ephemeral
regions to large active regions.

The longitude of flux emergence for each bipole is in principle
randomly drawn from a uniform distribution. However, this is modulated
by where previously emerged regions continue to exist: on the Sun,
bipolar regions have a strong tendency to emerge at locations where a
previous generation emerged and often still exists. This nesting
property is so pronounced that almost one in two active regions
emerges inside another \citep{Brouwer+Zwaan1990,1993SoPh..148...85H}. The
nesting as seen for solar regions is modeled as described by
\citet{schrijver2000}. In short, the probability of emergence per unit
area inside a
magnetic plage region is set to 22 times that outside of a plage
region \citep[following][]{1993SoPh..148...85H}. Other than this
hysteresis in flux emergence, no preferred longitudes are introduced.

The selected fluxes are distributed as flux concentrations over two
adjacent patches of opposite \replaced{polarity using observed
  area-flux properties for solar active regions (such that the average
  initial flux densities within these patches are
  $\sim 180$\,G).}{polarity. Work by \citet{ref254} revealed that
  mature active regions have a mean flux density of 100--150\,G
  (excluding spots and pores), regardless of the region's age or
  size. Here, a mean flux density is chosen of 180\,G (as in the
  original model by \citet{schrijver2000}) to accommodate a characteristic
  fraction of the flux contained in spots and pores.} The regions are
gradually introduced, with an equivalent rate of flux emergence of
$5\times 10^{21}$\,Mx/day \citep[based on data from][]{1993SoPh..148...85H}.

\added{The flux-transport model uses a point-source approximation for
  the magnetic field. These sources can be thought of as moving along
  a pattern of vertices that are defined by the evolving supergranular
  network. The properties of that cellular pattern and the
  characteristic displacement velocity along its vertices determine
  the typical mean-free path for the flux concentrations.  At each
  time step, the flux-transport model assumes that all flux within an
  area with a radius equal to the typical mean-free path length for
  the quiet-Sun network coagulate into a single concentration prior
  to the next time step. This concept was successfully tested for a
  range of solar activity levels.  The derivation of the coagulation
  radius $r_{\rm c}=4200$\,km is described in \textsection\,4 of
  \citet{schrijver2000}, based on the model by
  \citet{schrijver+etal1996e}.

  In reality, the supergranular and larger-scale flows that transport
  magnetic flux are, of course, not strictly confined to the lanes
  between supergranules, while moreover the supergranular cell pattern
  itself is evolving. Hence, the flux concentrations in the model are
  abstractions of patches within the photospheric field for which the
  diffusion approximation can be applied on scales beyond the
  coagulation radius. On scales below $r_{\rm c}$, granular convective
  motions interact with the magnetic flux through the laws of
  radiative magnetoconvection, reaching a quasi-equilibrium on a time
  scale of a few granular turnover time scales, well below the step
  time of 0.25\,day used in the model. Note that the diameter of the
  coagulation patch of $2r_{\rm c}= 8400$\,km is comparable to the
  box width of 9000 km used for the numerical models by
  \citet{2017A&A...605A..45N}, which is used in Sect.~\ref{sec:observations} to
  model the photospheric brightness. Hence,
  Sect.~\ref{sec:observations} uses the fluxes $\Phi_i$ in the point sources
  of the surface flux-dispersal model, converted into a mean flux
  density of $\Phi_i/(\pi r^2_{\rm c})$.}

The flux concentrations initiate a grid-free random walk
in which step lengths are a function of the absolute flux of the
concentrations, again following observed trends that show larger
concentrations to be less mobile, {\em i.e.}, apparently more able to
resist convective displacement.

These concentrations can collide to merge or 
(partially) cancel, and they can fragment, with collision cross sections and
fractionation probabilities depending on the flux contained, as
derived from 
solar observations.

Then, before starting on the next step, all concentrations are moved
subject to the large-scale differential-rotation and meridional-flow
profiles, after which a new set of bipolar regions is selected to
match the average flux-input rate, which itself is modulated following
the progression of an activity cycle shaped as the average
sunspot cycle.

I use the following parameter settings \citep[for details
see][]{schryver+title2001}: A time step of 6\,hr, a flux-dispersal
coefficient of $D=300$\,km/s$^2$ with a flux-dependent step size as
described by Eqs.~(A4) and~(A5) in \citet{schrijver2000}. The
differential-rotation rate is set to that from
\citet{Komm+Howard+Harvey1993c} and the meridional flow, tapered at
high latitudes, like that of \citet{vanballegooijen+etal98a}, both
as applied in the study by \citet{schryver+title2001}.  The half-life
time scale on which flux concentrations 'decay' (i.e., are randomly
removed from the simulation) as introduced and described by
\citet{schrijver+etal2002a} is set to 5\,yr. The model is set to use a fixed-amplitude
cycle with equal strengths at each sunspot maximum, set to reach its
peak 4\,yr after sunspot minimum, with a cycle-to-cycle overlap period
of 3\,yr \citep[see][]{schrijver+etal2002a}. The duration of the full magnetic
cycle is set to 21.9\,yr, and its associated flux-injection profile
in time and latitude is as in Eq.~(2) in \citet{schryver+title2001}.

\subsection{Model runs}\label{sec:runs}
The model is run for five different sets of only two free
parameters. The first, ${\cal A}$, quantifies the amplitude of the
stellar cycle; it is simply a multiplier on the frequency at which
bipolar regions are inserted compared to a typical solar cycle, so
that ${\cal A}=1$ is used for a run to mimic the Sun \added{(calibrated
to Cycle~21)}, while ${\cal A}>1$ signifies a more-active star with a
more pronounced cycle. The second parameter, $\Phi_{\rm max}$, is a threshold
flux in bipolar regions at which the power-law frequency spectrum of
fluxes is truncated, as discussed below. The parameters and results of
the five runs are summarized in Table~\ref{tab:modulations}.

A star like the present-day Sun is simulated as in the model
developed by \citet{schrijver2000}, here referred to as ${\cal M}({\cal A}=1)$.  For this
baseline solar model, the bipole-emergence rate is calibrated to
the solar cycle, i.e. with a cycle-strength multiplier as in
\citet{schryver+title2001} of ${\cal A}=1$. 

For stars like Kepler-17 and WASP-36, of solar spectral type but with
rotation periods around 12\,days, a coronal soft X-ray flux density is
expected that is about 10 times higher than that of the Sun
\citep[{\em e.g.},][]{patten+simon96,2018AnA...618A..48M}. As the soft
X-ray flux density scales roughly linearly with the surface magnetic
flux density \citep[{\em e.g.},][]{schrijver+title2004} the mean
surface magnetic flux density, $\langle |fB| \rangle$ (for filling
factor $f$ and intrinsic photospheric field strength $B$), should be
roughly an order of magnitude higher than for run
${\cal M}({\cal A}=1)$, here particularly for the facular fields
because spot fields appear to contribute little to the coronal
brightness, so roughly $\sim 150$\,G. Reaching such levels of surface
magnetic activity requires significant enhancements in the
active-region injection frequency or mean active-region size, or both.

\added{This estimate of the average surface magnetic flux density
$\langle |fB| \rangle \sim 150$\,G for stars of about half the solar
rotation period is subject to a considerable uncertainty range. One
discussion of this (with associated references) can be found in
\citet{2014MNRAS.441.2361V}: using the five power-law scalings from
various sources in the literature that are listed in their Table~3
based on Zeeman broadening to quantify the surface magnetic field, one
infers $\langle |fB| \rangle=$40--160\,G starting from a solar value
from run ${\cal M}({\cal A}=1)$ of $\sim 15$\,G. A value of $\sim 170$\,G
results from a calibration from Rossby number via Zeeman-Doppler
imaging to an equivalent magnetic flux density for a Zeeman broadening
signal using Figure~1 and Eq.~(2) in
\citet{2019ApJ...876..118S}. Thus, the value of $\sim 150$\,G
estimated above seems reasonable, although a substantial uncertainty
should be allowed for.}

A much more active star, but with otherwise solar parameters, is
simulated using ${\cal A}=30$ in model ${\cal M}({\cal A}=30)$.  The
total flux on the stellar surface in the latter run peaks at about 7
times that of ${\cal M}({\cal A}=1)$: inserting more active regions with a
given size spectrum leads to more flux cancellations, so that the total
flux on the surface increases more slowly than linearly with the cycle
strength \citep[see, e.g.,][for a
discussion]{schrijver2000}. Whereas it reaches a value of
$\langle |fB| \rangle$ commensurate with that of a star with
$P_{\rm rot}\sim 12$\,days, the rotational amplitude is too low (see
Sect.~\ref{sec:rotmod}).

To increase the rotational modulation, larger regions could be
introduced. This also strengthens the impact of activity nests, the
tendency for active regions to emerge within existing other regions or
in sites of an earlier such emergence.  Thus, another run,
${\cal M}({\cal A}=30)^+$, also with ${\cal A}=30$, is created but
with a maximum active-region flux per polarity of $\Phi_{\rm max}$
raised from $1.5\times 10^{22}$\,Mx \citep[as used in][]{schrijver2000} to
$3\times 10^{23}$\,Mx.  The latter value allows regions as large as that
expected to power the Carrington-Hodgson flare to emerge \citep[based
on a flux estimate by][]{2013A&A...549A..66A}. Such very large regions
are rare on the Sun \added{\citep[the largest region observed to date,
  which occurred in April 1947, is estimated to have had a flux at that level,
  see][]{taylor1989,2012JGRA..117.8103S}}, hence the choice of
$\Phi_{\rm max}=1.5\times 10^{22}$\,Mx for the solar activity level in run
${\cal M}({\cal A}=1)$, but the occurrence of very large flares on
more-active stars suggests that larger regions may be more common on
such stars \citep[see, {\em
  e.g.},][]{2012Natur.485..478M,2013arXiv1308.1480S,2017PASJ...69...12N}.

The model runs described above suggested that stars like 
Kepler-17 and WASP-36 may be approximated by a run of intermediate ${\cal A}$ and
$\Phi_{\rm max}$, with parameters set as in run ${\cal M}({\cal A}=10)^+$.
For comparison purposes, also a run
${\cal M}({\cal A}=1)^+$ is executed for a star of solar activity, but
allowing for large active regions, setting
$\Phi_{\rm max}=3\times 10^{23}$\,Mx.

Note that all runs for an active Sun-like star with ${\cal A}=30$ have
the same total amount of flux emerging per unit time in the range from
$6\times 10^{18}$\,Mx to $1.5\times 10^{22}$\,Mx per polarity. The larger range
of the power-law probability distribution for regions to be inserted
into the stellar photosphere in runs with elevated $\Phi_{\rm max}$
introduces additional large regions, leading to more flux on the
stellar surface, while maintaining the frequencies of the more
abundant smaller regions to match the shape of the size spectrum
derived from solar observations (while allowing an overall
multiplicative factor). Given the properties of the active-region
frequency distribution, the rate of flux input for
$\Phi_{\rm max}=3\,10^{23}$\,Mx is increased by a factor of 1.7 relative to
$\Phi_{\rm max}=1.5\,10^{22}$\,Mx. Larger regions disperse more
slowly, which also leads to increased persistence of successive
generations of flux emergence within them (in active-region nests), so
that the total amount of flux in the photosphere increases more than
the rate at which flux input increases if the spectrum is extended to
higher fluxes (see
Table~\ref{tab:modulations}).

To bootstrap the simulation runs relatively quickly into a state in
which the polar caps are less sensitive to the initial state (an empty
stellar surface), for all runs except ${\cal M}({\cal A}=30)^+$, half
of all flux concentrations is randomly removed after the completion of
the first 11-yr sunspot cycle \citep[using the same procedure as
in][]{schryver+title2001}. For these runs, the data shown here are
taken from the third sunspot cycle, i.e., covering the years 22 to 33
for the simulations. Any effects of the initial state or the removal
of half of all concentrations after the first 11 years are moreover
dampened by the half-life decay time scale of flux concentrations of
5\,yr. For run ${\cal M}({\cal A}=30)^+$, when the code is tracking
over $2\times 10^5$ elements near cycle maximum, the $n^2$ nature of the
algorithm testing for collisions between flux concentrations leads to
much longer computation time; as only the high-latitude fields require
a multi-decade run, while these have little impact on the transit
residuals being studied, for this run data are shown in the same cycle
phases as for the others, but from the first rather than third sunspot
cycle.

\begin{figure*}[th]
\centerline{(a) ${\cal M}({\cal A}=1)$, $\Phi_{\rm max}=1.5\,10^{22}$\,Mx: \hskip 0.0cm 3870\,\AA\ \hskip 2.5cm 6010\,\AA\
  \hskip 3.6cm 15975\,\AA}
  
\includegraphics[clip=true, width=9.0cm]{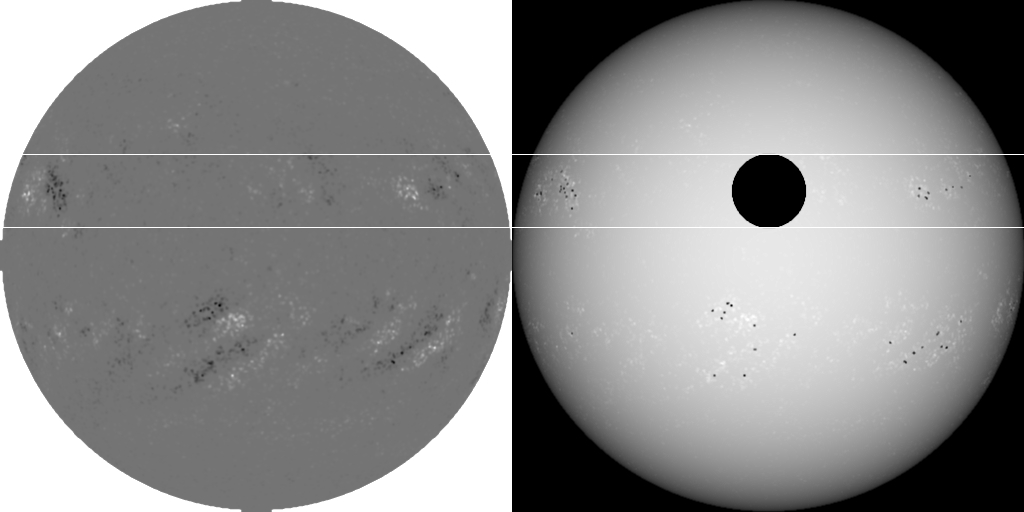}\includegraphics[trim=512 0 0 0, clip=true, width=4.5cm]{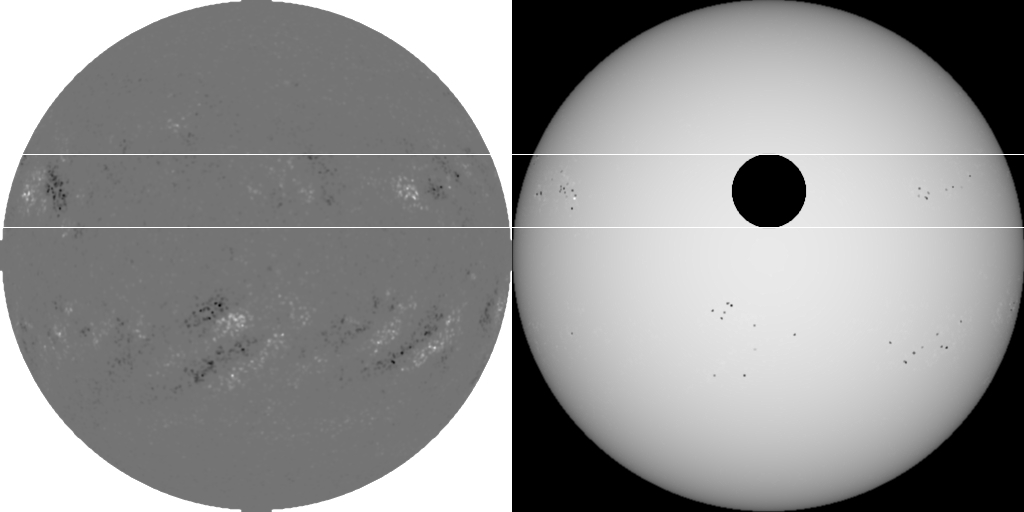}\includegraphics[trim=512 0 0 0, clip=true, width=4.5cm]{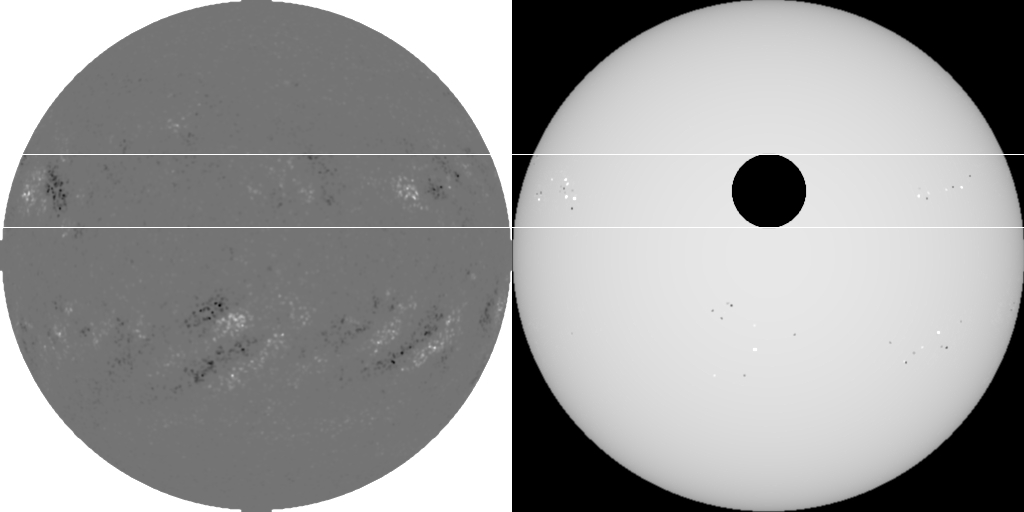}

\centerline{(b) ${\cal M}({\cal A}=1)^+$, $\Phi_{\rm max}=3.0\,10^{23}$\,Mx:  \hskip 0.0cm  3870\,\AA\ \hskip 2.5cm 6010\,\AA\
  \hskip 3.6cm 15975\,\AA}

\includegraphics[clip=true, width=9.0cm]{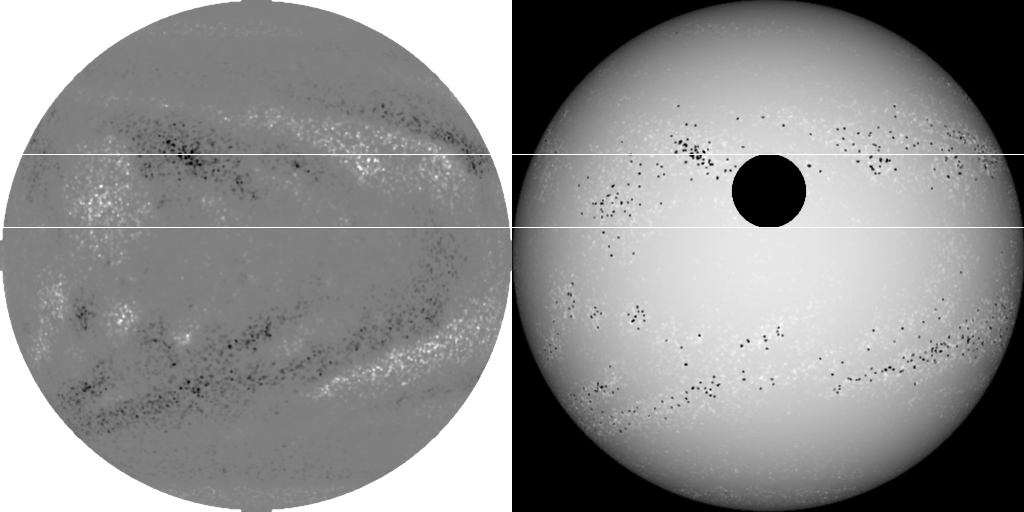}\includegraphics[trim=512 0 0 0, clip=true, width=4.5cm]{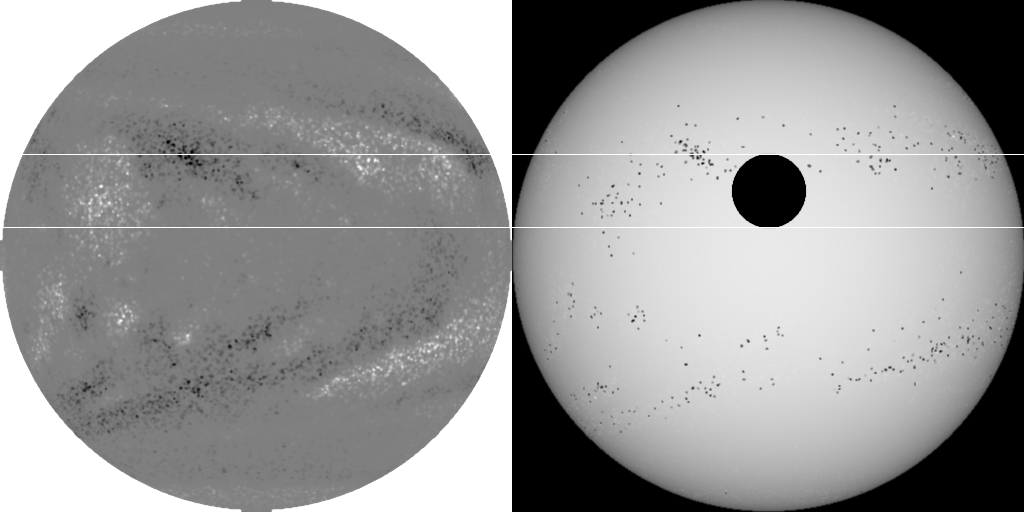}\includegraphics[trim=512 0 0 0, clip=true, width=4.5cm]{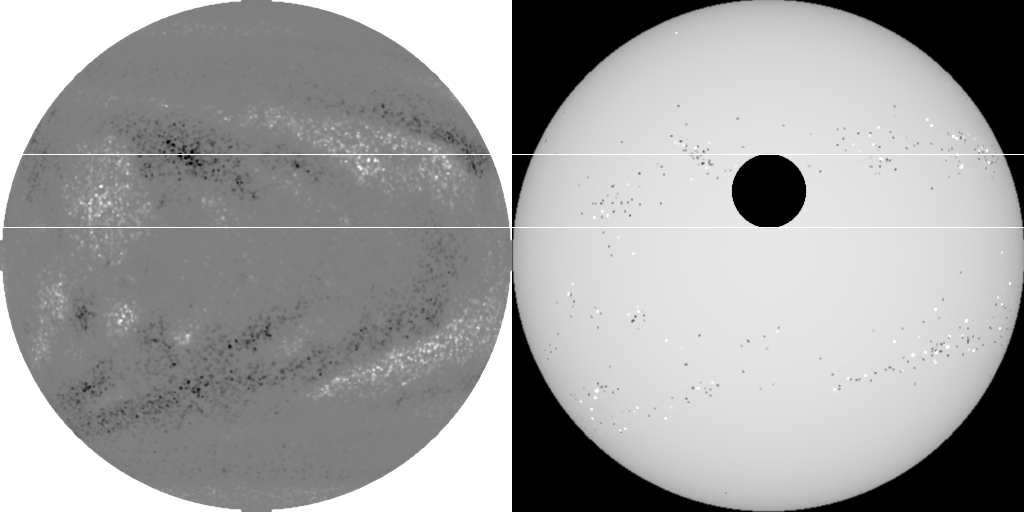}

\centerline{(c) ${\cal M}({\cal A}=10)^+$, $\Phi_{\rm max}=3.0\,10^{23}$\,Mx:  \hskip 0.0cm  3870\,\AA\ \hskip 2.5cm 6010\,\AA\
  \hskip 3.6cm 15975\,\AA}

\includegraphics[clip=true, width=9.0cm]{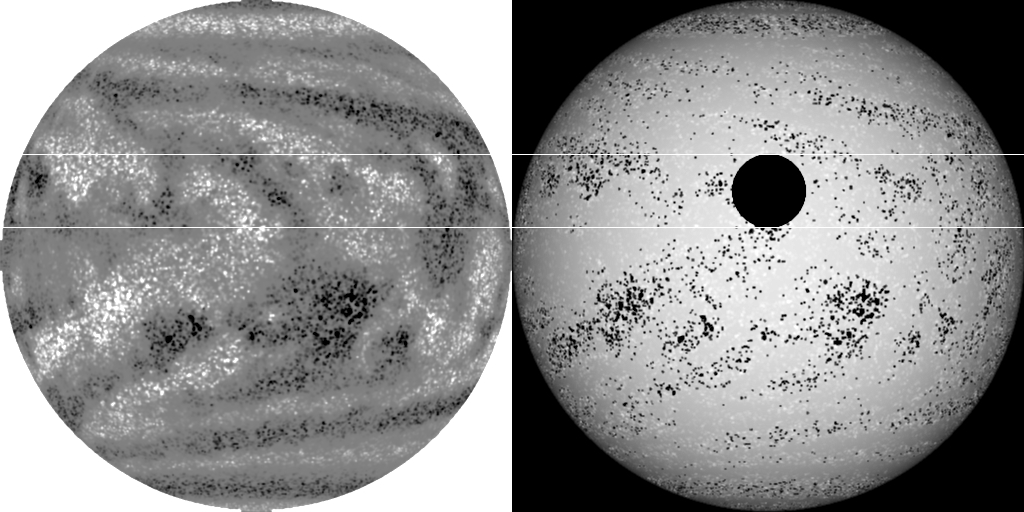}\includegraphics[trim=512 0 0 0, clip=true, width=4.5cm]{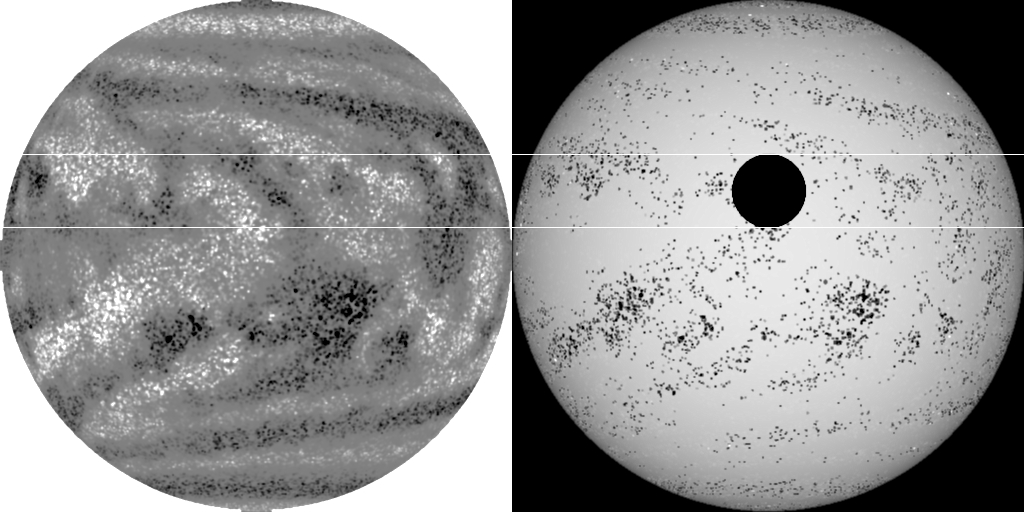}\includegraphics[trim=512 0 0 0, clip=true, width=4.5cm]{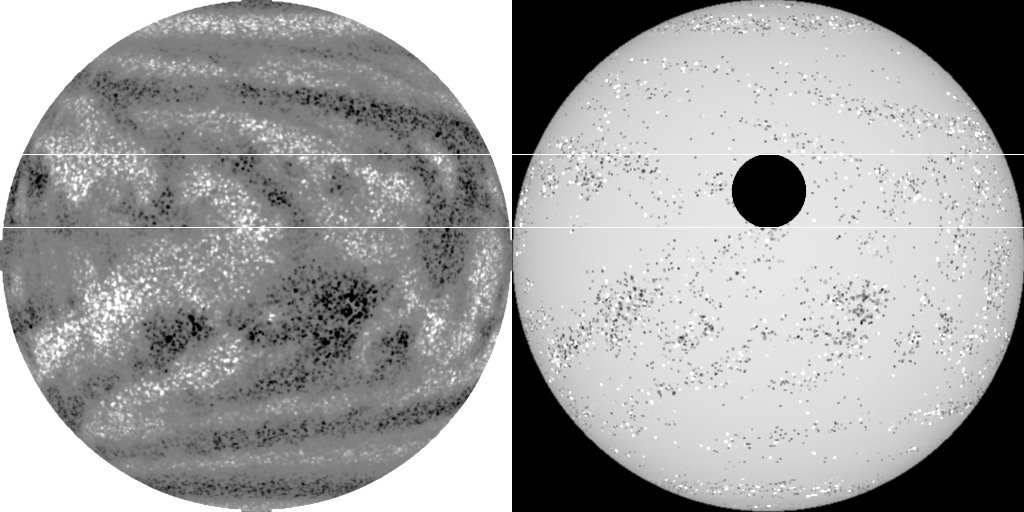}

\centerline{(d) ${\cal M}({\cal A}=30)^+$, $\Phi_{\rm max}=3.0\,10^{23}$\,Mx:  \hskip 0.0cm  3870\,\AA\ \hskip 2.5cm 6010\,\AA\
  \hskip 3.6cm 15975\,\AA}

\includegraphics[clip=true, width=9.0cm]{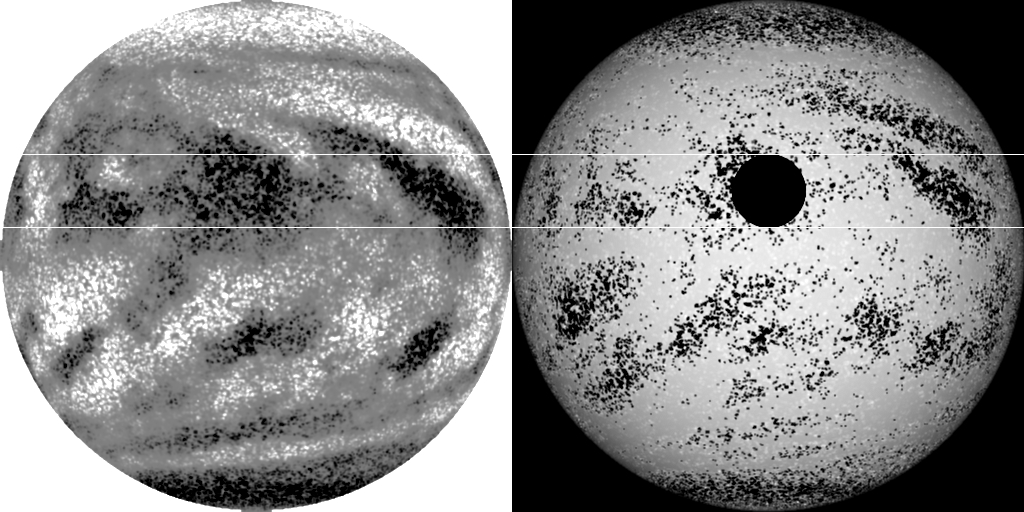}\includegraphics[trim=512 0 0 0, clip=true, width=4.5cm]{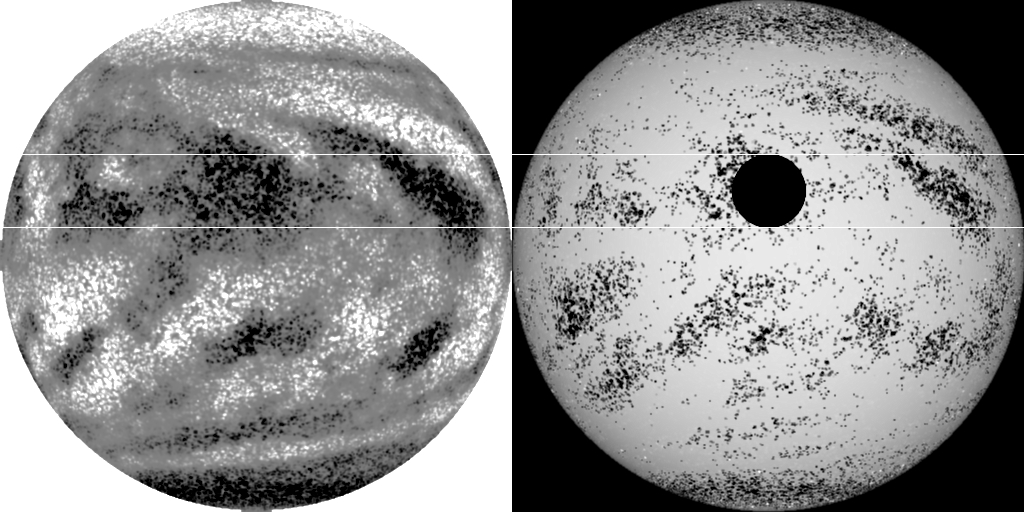}\includegraphics[trim=512 0 0 0, clip=true, width=4.5cm]{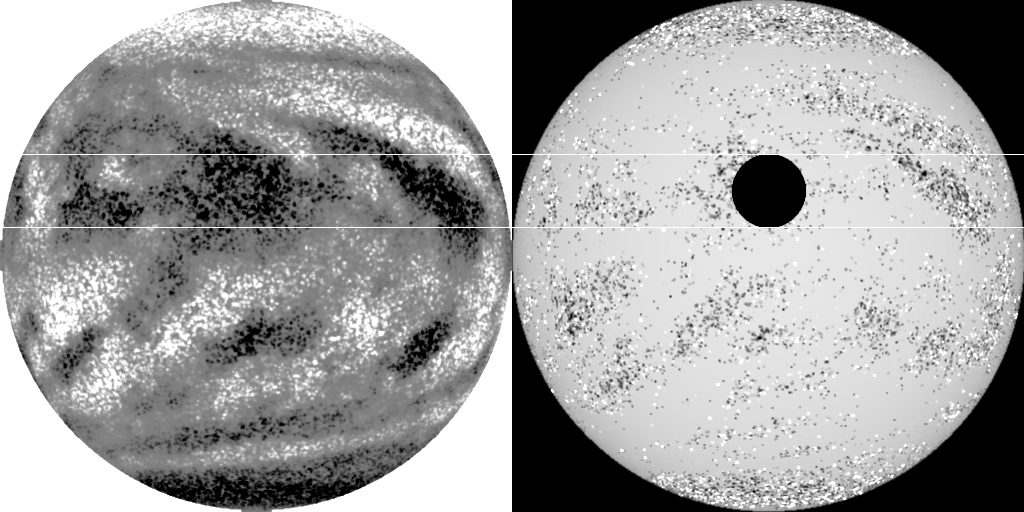}
\caption{\label{fig:example79} Simulated magnetograms (left panels;
  gray scale saturates at $\pm 1.35$\,kG) and intensity images for
  3870\,\AA, 6010\,\AA, and 15975\,\AA, respectively, for a Sun-like
  star, approximately 5.5\,yr after the minimum in its solar-like
  sunspot cycle. The intensity contrast relative to the quiet
  photosphere has been doubled to better show the faculae. The cycle
  amplitude ${\cal A}$ and the maximum flux $\Phi_{\rm max}$ per
  polarity in the largest active regions (see Sect.~\ref{sec:runs} for
  their definition) are indicated above the
  magnetograms. \added{The transit path and the exoplanet size used in
    Sect.~\ref{sec:transits} are shown by white lines and the black
    disks.}}
\end{figure*}

Sample simulation results for the magnetic field distributions across
the stellar surface are shown in
Figures~\ref{fig:example79}a-\ref{fig:example79}d. As the focus of this
study is the signal from exoplanetary transits, and as most
planetary systems appear to have their normal vector rather well aligned with
the stellar rotation axis, the perspective for visualizations of 
these simulations is chosen to lie in the plane of the stellar rotational
equator. 

\subsection{Modeling photospheric appearances}\label{sec:observations}
Estimation of the brightness of facular patches on the faces of stars
requires knowledge of their contrast as a function of viewing angle (or
relative distance from disk center), wavelength, and corresponding
characteristic absolute magnetic flux densities.  Here, disk
appearance, rotational modulation, and transit signals are simulated
for three wavelengths for which \citet{2017A&A...605A..45N} show
facular contrast versus limb distance (their Fig.~8) and activity
level: 3870\,\AA, 6010\,\AA, and 15975\,\AA. Their values are based on
magnetoconvective calculations with radiative transfer using the MURaM
code, validated successfully against solar observations.\footnote{The
  applicability of these models to stars substantially more active
  than the Sun is supported by, e.g., the observations by 
  \citet{2012ApJ...745...25L} who conclude that the thermal structure
  and heating rates inferred for active stars are comparable to those
  of the brightest solar faculae \citep[see][for a broader
  discussion]{2019LNP...955.....L}.} For values of
$\mu$ between 0 and 0.2, close to the limb, for which they show no
results, a smooth transition to zero contrast at the limb is imposed.

The background limb darkening of the quiet photosphere is set to that
measured by \citet{1994SoPh..153...91N}, except for the 15975\,\AA\
channel for which I use the polynomial fit to the data presented in
Figure~4 by \citet{2017A&A...605A..45N}.

To estimate the average magnetic flux density to be used to compute
facular brightenings, I assume that the flux in a model concentration
is spread out over an area with a radius equal to the coagulation radius of
$r_{\rm c}=4200$\,km: concentrations that approach each other closer
than that in the model are considered to merge. The flux threshold at which
bright faculae transition to dark pores or spots is set to
$\Phi_{\rm b}=3\times 10^{20}$\,Mx \citep[see Table~4.1
in][]{schrijver+zwaan99}.

For spots, the average intrinsic field strength over the combined area
of umbra and penumbra is set to 1.35\,kG based on the study by
\citet{1993AnA...267..287S}, who note that this value is similar to
the field strength characteristic of smaller magnetic features; this
value is therefore used as intrinsic field strength in the visualizations here
for all flux concentrations. For
the intensity contrast averaged over an entire pore or spot, I assume
an average intensity contrast derived from the ratio of blackbody
temperatures of the spot and photosphere, $T_{\rm spot}$ and
$T_{\rm phot}$, respectively, corresponding to the wavelength
$\lambda$ considered:
$c_{\rm s}(\lambda)= \left (\exp(hck/(\lambda T_{\rm phot}))-1\right)/\left (
  \exp(hck/(\lambda T_{\rm spot}))-1\right)$, for $h$ the Planck
constant, $c$ the speed of light in vacuum, and $k$ the Boltzmann
constant. Radiative transfer effects for different wavelengths are
ignored in the present approximation.

For the Sun-like star simulated here, the photospheric temperature is
set to $T_{\rm phot}=5780$\,K.  For the effective temperature of the
spot area, including umbra and penumbra, I use $T_{\rm spot}=5150$\,K
as inferred from solar irradiance modeling by
\replaced{\citet{1998A&A...335..709F}. With these
  values,}{\citet[][see also, {\em e.g.},
  \citet{2001JASTP..63.1479F}]{1998A&A...335..709F}.  More recent
  modeling may include umbral and penumbral components separately,
  such as in the study by \citet{2008A&A...486..311U} which uses a
  penumbral temperature of 5150\,K and an umbral temperature of
  4500\,K. However, as the penumbral area dominates umbral areas by
  factors of typically 3:1 \citep{1998A&A...335..709F} to 5:1
  \citep{1993A&A...274..521M}, using a single, largely penumbral
  temperature as characteristic for the overall spot output suffices
  for the present purpose. With these values for $T_{\rm phot}$ and
  $T_{\rm spot}$,} the spot brightness relative to the photosphere at
3870\,\AA, 6010\,\AA, and 1.6\,$\mu$m is 0.45, 0.60, and 0.79,
respectively.  The same relative limb-darkening curve is assumed as
for the quiet photosphere at the wavelength under consideration.

Each flux concentration is mapped into the image pixel corresponding
to its central location.  For those with $|\Phi_i| \le \Phi_b$, the
effective radius $A_{\rm eff}$ is assumed to be $r_{\rm c}$.  For
concentrations with $|\Phi_i| > \Phi_b$, an area of
$A_\Phi=|\Phi_i|/B_{\rm phot}$, with $B_{\rm phot}=1.35$\,kG is assumed to
hold for the average field strength over the combined umbral and
penumbral areas.  For the concentrations for which $A_\Phi$ exceeds
the unprojected surface area mapping into an image pixel, the area
is approximated as a foreshortened circular disk (pixelation effects
are corrected for by matching the total intensity in a disk to the area
mapped onto an image array). For fractional pixels, and
for concentrations for which the area $A_{\rm eff}$ is smaller than
the solar surface area under an image pixel, the intensity in that
pixel is modified using the fraction of the pixel area covered by the
magnetic concentration, assuming the intensity of the complement of
that pixel area is unaffected.

Simulated magnetograms are made by mapping the fluxes $\Phi_i$ of the
flux concentrations onto an
image, spread out over disks with radius $A_\Phi$, and subsequently
modified with a multiplicative factor for each pixel equivalent to
assuming that the magnetic field is normal to the local surface.

Sample simulation results for the intensity images at three different
wavelengths are shown in the right-hand three panels of
Figures~\ref{fig:example79}a-\ref{fig:example79}d (with doubled contrast
to better show the faculae).
The images in these figures were smoothed with a
Gaussian with a FWHM value of 2~pixels.
The images show that both the facular contrast and the spot
contrast are largest for the shortest wavelengths, as expected from
the facular contrasts and the wavelength-dependent contrast
$c_{\rm s}(\lambda)$. And, also as expected, the faculae stand out most
clearly toward the limb while spots and pores are most readily
visible toward the central regions of the disks.

One property that jumps out in particular in the intensity images of
Figures~\ref{fig:example79}c--\ref{fig:example79}d is the abundance of
dark features with diameters of order 8000\,km. This is a consequence
of the numerical prescription for the fractionation probability of
flux concentration. This has a local minimum (and thus a corresponding
maximum in the lifetime) around $10^{21}$\,Mx. Consequently, clusters
of such solar-like spots develop in the model. While this
fractionation probability approximates the behavior of most sunspots
well, it is as yet unknown how spots and spot clusters behave in very
active stars or extremely large bipolar regions (if these indeed
exist). \added{Numerical experiments with radiative magnetoconvection
  by \citet{beeckmuram2015a,beeckmuram2015b}, however, suggest that in
  the presence of an increasingly strong field, (micro-)pores are a
  naturaly occuring phenomenon for cool main-sequence stars when the
  mean flux density exceeds a few hundred Mx/cm$^2$.}  This may or may
not be what happens on a very active star, but even if it does not happen, compact
clusters of such spots in the model run here may effectively describe
the appearance of any large spot in real life from the point of view
of rotational modulation or transit signals for planets of a size
comparable to Jupiter as simulated here. Note that the images shown in
Figures~\ref{fig:example79}a--\ref{fig:example79}d have a best
effective resolution of 5500\,km at disk center (given the diameter of
the rendered images of 512 pixels subjected to a Gaussian smoothing
with FWHM of 2 pixels), to be compared to $r_{\rm c}=4200$\,km.

Another feature that stands out in
Figures~\ref{fig:example79}c--\ref{fig:example79}d is the multitude of
spots and pores that persist in the poleward arcs formed by decaying
active regions, and also in flux dispersal toward the equator.  In
the models, these dark features are the result of the frequent
collisions and temporary mergers between flux concentrations in
regions of high average flux density. Such high flux concentrations
on the Sun would correspond to pores and small spots.

\begin{table*}
  \caption{\label{tab:modulations} Model properties and modulation
    estimates. The table lists the following for a single,
    arbitrarily-chosen step in the flux-transport model for an
    activity phase about 1.5\,yr after cycle maximum, following the
    run identifier, which includes the cycle-strength factor
    ${\cal A}$, and the maximum active-region flux per polarity
    $\Phi_{\rm max}$: \replaced{surface}{globally}-averaged absolute
    magnetic flux density $\langle |fB| \rangle$ of all flux elements
    and $\langle |fB| \rangle_{\rm fac}$ for the faculae (with fluxes
    below $3\,10^{20}$\,Mx); \replaced{surface}{globally}-averaged
    areal filling factor $f_{\rm d}$ of dark pores and spots assuming
    an intrinsic field strength of 1.35\,kG; the contrast
    $\delta_\ast$ of the total unocculted brightness of the active
    star and the unocculted inactive star from the perspective shown
    in Figure~\ref{fig:example79}; peak-to-trough rotational modulation,
    and --~in the final set of three columns~-- peak-to-trough amplitude
    $\Delta(\lambda)$ for the transit residuals, both as shown in
    Figures~\ref{fig:example79}a--\ref{fig:example79}d. Note that
    $\delta_\ast(\lambda)$ and $\Delta(\lambda)$ are for rough
    intercomparison only as they are computed for a single time step
    in the flux-dispersal model and for a transit across a single
    viewing angle of the stellar sphere.}
\begin{tabular}{ll|cr|lll|lll|rrr}
\hline 
Run  &$\Phi_{\rm max}$& $\langle |fB| \rangle$, & $f_{\rm d}$ &
            \multicolumn{3}{c|}{Net Intensity}  & \multicolumn{3}{c|}{Rotation}  &
            \multicolumn{3}{c}{Transit Residual} \\
&($10^{22}$& $\langle |fB| \rangle_{\rm fac}$ & (\%) &
            \multicolumn{3}{c|}{Change  $\delta_\ast$ (\%)}  &
            \multicolumn{3}{c|}{Modulation (\%)}      &
            \multicolumn{3}{c}{Ampl. $\Delta(\lambda)$ (ppm)} \\
& Mx)& (G)& \null &3870 & 6010 & 15975 & 3870 & 6010 & 15975 & 3870 & 6010 & 15975 \\
\hline 
${\cal M}({\cal A}=1)$        & 1.5     &16, 15&0.06&$+0.24$&$+0.04$&$+0.014$&0.06&0.014&0.010&180&40&40\\ 
${\cal M}({\cal A}=1)^+$  & 30      & 60, 51&0.65&$+0.55$&$-0.01$&$-0.00$&0.38&0.31&0.12&420&180&70\\ 
${\cal M}({\cal A}=10)^+$  & 30      &212, 132&5.9&$-0.97$&$-1.60$&$-0.40$&0.90&0.64&0.25&980&520&300\\ 
${\cal M}({\cal A}=30)$   & 1.5      &135, 100&2.6&$+0.26$&$-0.54$&$-0.12$&0.30&0.22&0.08&1100&540&190\\ 
${\cal M}({\cal A}=30)^+$  & 30      &479, 175&22.&$-6.1$&$-5.3$&$-1.3$&2.9&2.0&0.7&6300&3800&1400\\ 
\hline 
  \end{tabular}
\end{table*}

\begin{figure}
  \centerline{\includegraphics[width=7.0cm]{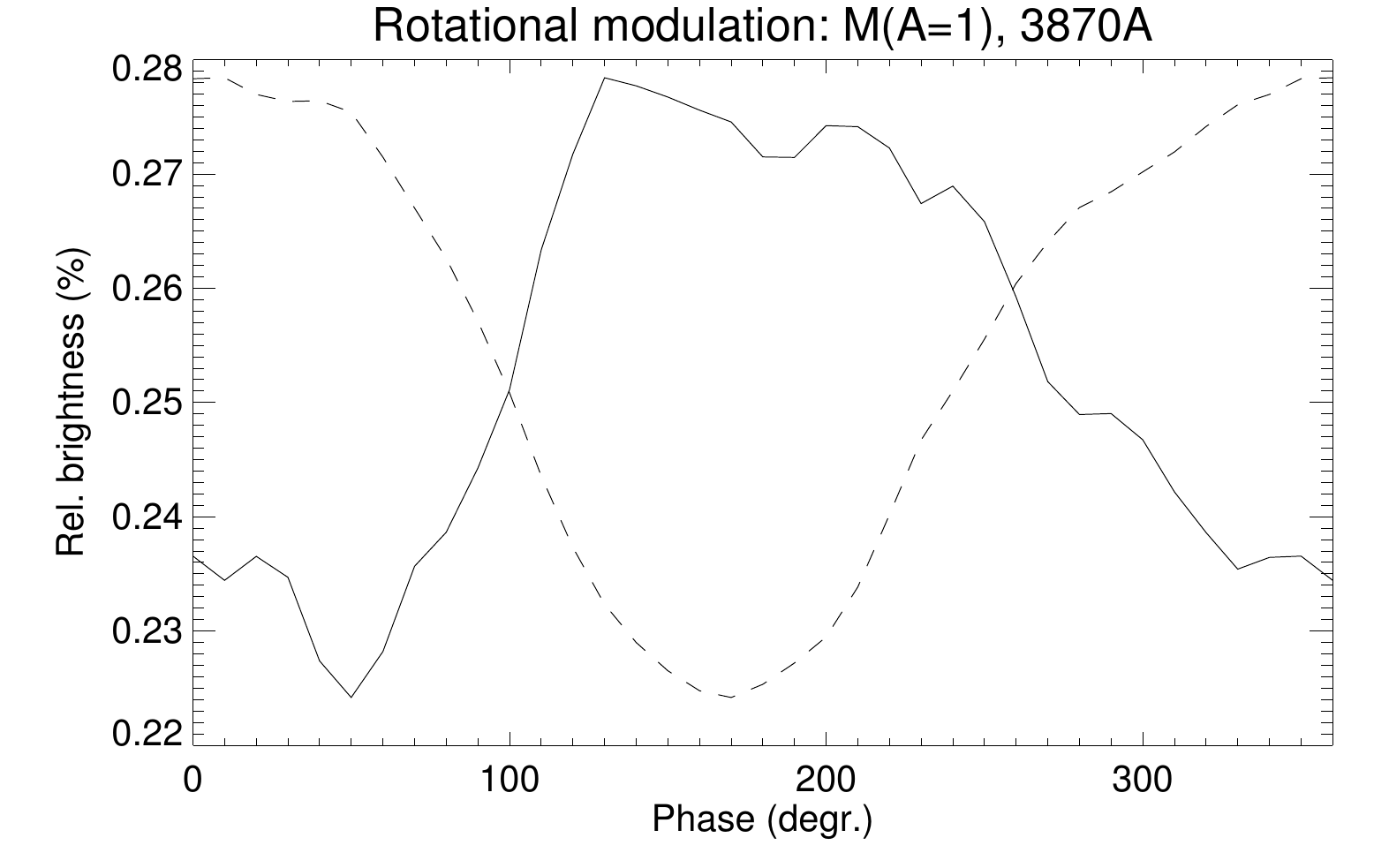}}
  \caption{\label{fig:rotmod79} Solid line: rotation modulation for
    3780\,\AA\ for run ${\cal M}({\cal A}=1)$, \replaced{measured}{as viewed from
      the rotational equator, and expressed} as a
    difference compared to an inactive but limb-darkened photosphere
    such that phase angle $0^\circ$ (corresponding to the images in
    Fig.~\ref{fig:example79}a) shows the offset $\delta_\ast$ from
    Table~\ref{tab:modulations}. Dashed line: total flux on the
    observer-facing side of the star, rescaled to the same amplitude as the
    intensity but inverted in sign.}

  \centerline{\includegraphics[width=7.0cm]{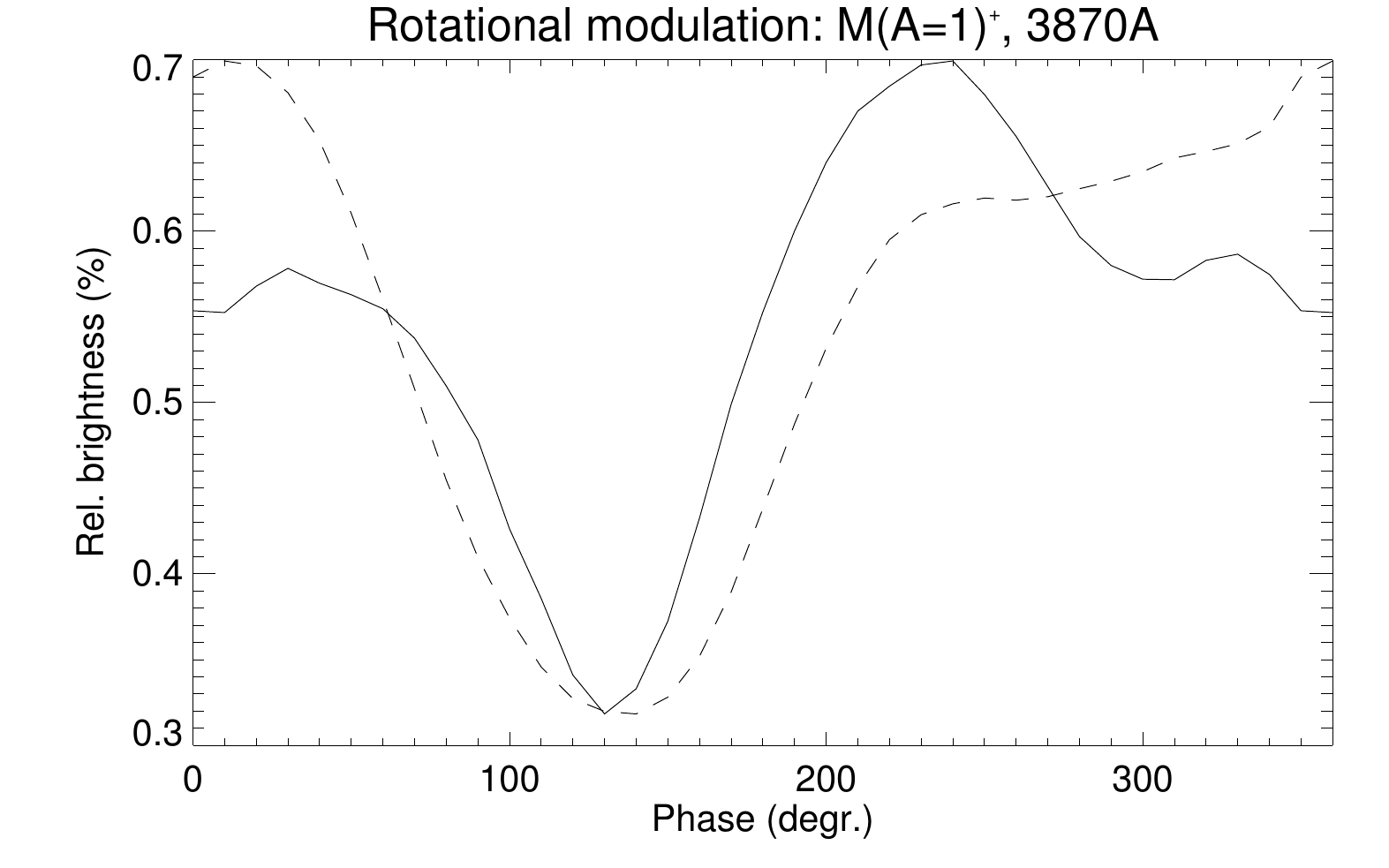}}
  \caption{\label{fig:rotmod82} As Figure.~\ref{fig:rotmod79} for ${\cal M}({\cal A}=1)^+$. }

  \centerline{\includegraphics[width=7.0cm]{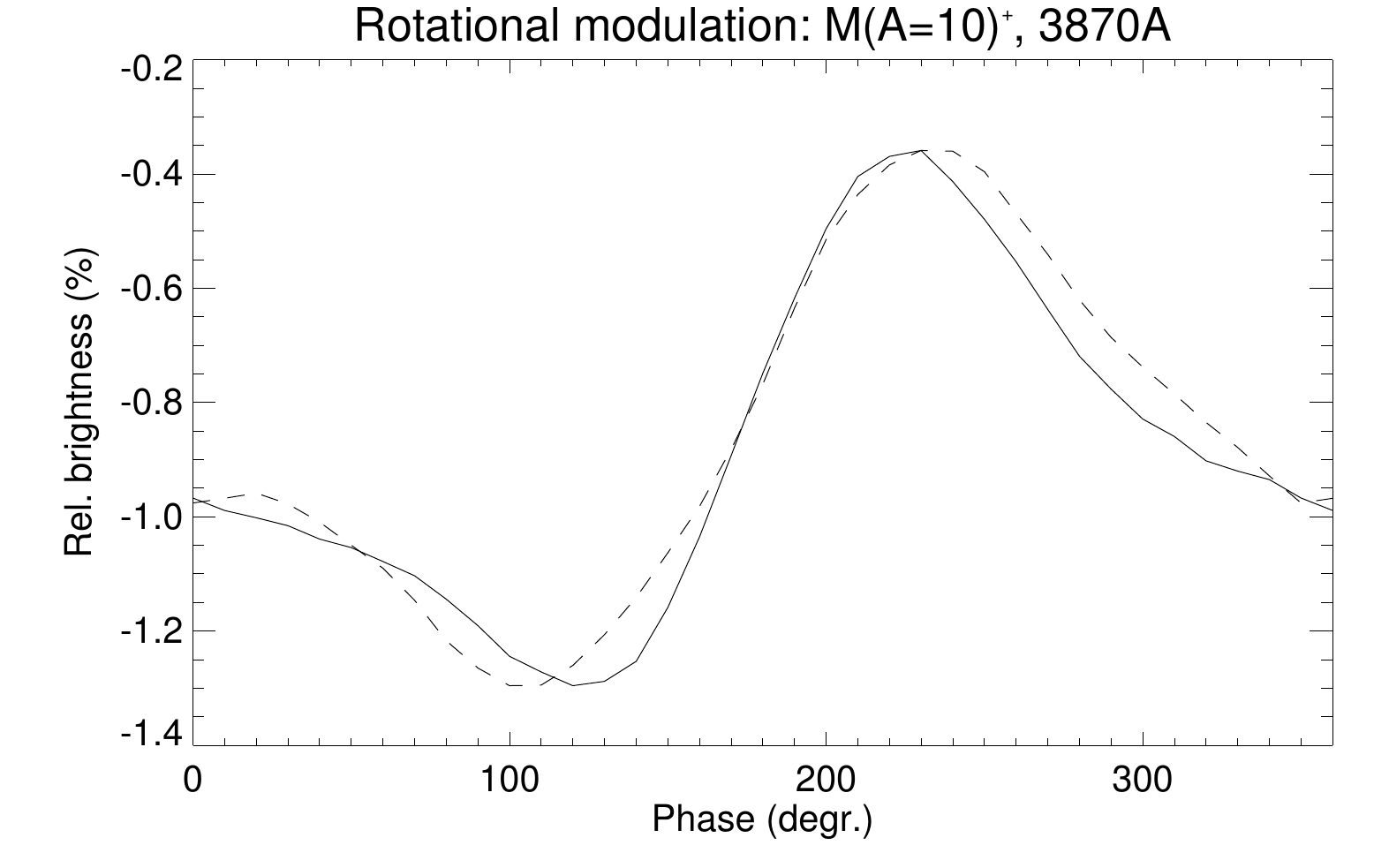}}
  \caption{\label{fig:rotmod83} As Figure.~\ref{fig:rotmod79} for ${\cal M}({\cal A}=10)^+$. }

  \centerline{\includegraphics[width=7.0cm]{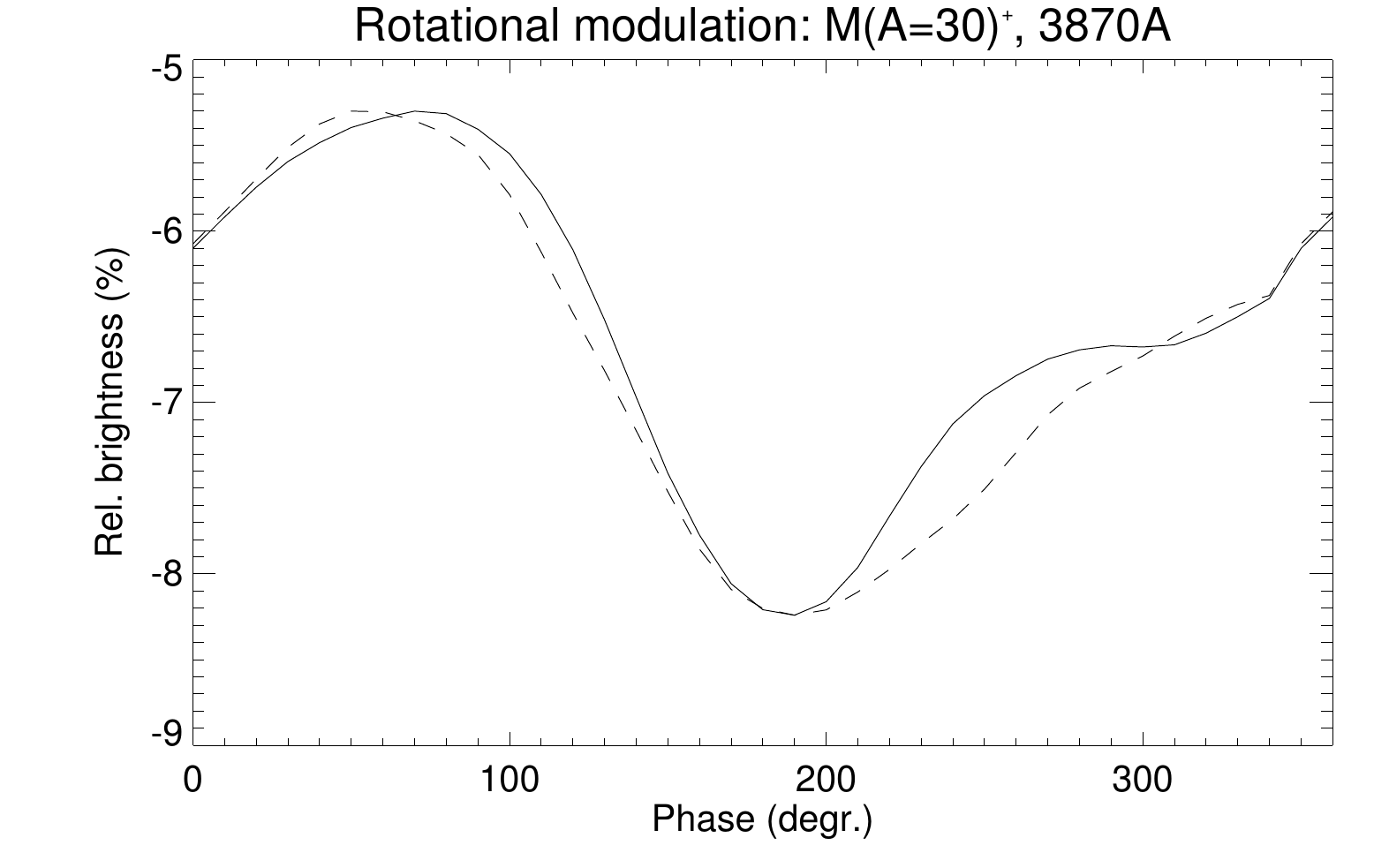}}
  \caption{\label{fig:rotmod80} As Figure.~\ref{fig:rotmod79} for ${\cal M}({\cal A}=30)^+$. }
\end{figure}
\section{Out-of-transit brightness and rotational
  modulation}\label{sec:rotmod}
This section reviews how the modeled stars would be viewed in terms of
brightness and rotational modulation in the absence of exoplanetary
transits. Only a single realization is randomly selected for this
study (5.5\,yr after a minimum, and 1.5\,yr after the maximum in the
simulated cycle), which focuses on trends in its testing of the solar paradigm
against selected observations.

\subsection{Net brightness}
First to discuss is the average relative offset
$\delta_\ast=(I_\ast(\infty,\lambda)-I_{\rm q}(\lambda))/I_{\rm q}(\lambda)$,
where $I_\ast(\infty,\lambda)$ is the brightness of the active star out
of transit and $I_{\rm q}(\lambda)$ is that of the quiet-star (here
quiet-Sun) photosphere. Values for $\delta_\ast$ are listed in
Table~\ref{tab:modulations}.  Note that these values should be
compared across wavelength for a given set; substantial statistical
fluctuations make these values less suitable for comparison between
runs, although the trends are likely to stand. These values show that
the tradeoff between bright faculae and dark spots is such that the
overall brightness of a star with Sun-like activity (${\cal M}({\cal A}=1)$)
is increased during cycle phases of enhanced activity relative to the
reference quiet Sun. The averaged contrast over the three wavelengths,
used as a rough proxy for the bolometric value, is $\sim 0.1$\%. This
value compares well with the observed change in total solar irradiance
over the cycle
\citep[{\em e.g.},][]{2016JSWSC...6A..18F,2016JSWSC...6A..30K}.

For the simulations of the more-active stars, the faculae play a
relatively weaker role than starspots, resulting in a dimming
of the overall stellar brightness. For ${\cal M}({\cal A}=1)^+$ this
results in a near balance between spot darkening and facular
brightening for the longer wavelengths, while for the most active star,
simulated sunspot darkening outweighs facular contributions at all
three model wavelengths. This trend is consistent with the empirical
results of \citet{lockwood+etal97} and \citet[][see also \citet{2017ApJ...851..116M}]{2018ApJ...855...75R}. In
their long-term monitoring of a set of cool stars, they note that
more-active stars tend to dim as chromospheric activity increases due
to dynamo variability, suggesting a dominant role of starspots over
faculae. In contrast, for many less-active stars --~including the
Sun~-- photospheric brightness increases with increasing activity,
which is interpreted to be owing to a stronger role of faculae relative to that of
spots, although the latter can cause dips in the brightness when
crossing near central meridian due to solar rotation.

\added{Figure\,15 by \citet{2018ApJ...855...75R} suggests that spots
  and pores dominate in irradiance variations beyond a chromospheric
  activity level of $\log(R^\prime_{\rm H+K})\approx -4.75$, compared
  to a characteristic solar value of around $-4.9$.
  Converting the level of
  $\log(R^\prime_{\rm H+K})\approx -4.75$ into an $S$-index value
  \citep{1984ApJ...279..763N}, and applying Eqs.~(1) and~(2) in
  \citet{2014A&A...569A..38S} yields a corresponding spot/pore filling
  factor of 1.2\%\ and a facular filling factor of 6.8\%, which with
  $B=1.35$\,kG yields $\langle |fB| \rangle \approx 100$\,G for
  spots/pores and faculae combined.  This appears compatible with the
  summary results in Table~2.}

\subsection{Rotational modulation}
To compute the rotational modulation as shown by the solid curves in
Figures~\ref{fig:rotmod79}--\ref{fig:rotmod80}, the perspective of the
rendered photospheres is changed, but the magnetic field is held
frozen in time (phase angle $0^\circ$ corresponds to the images shown
in Figures~\ref{fig:example79}a--\ref{fig:example79}d). Dashed curves in
these figures show the total magnetic flux on the observer-facing side of the star,
rescaled to the same range for comparison, and inverted in sign
because more flux is generally due to more active regions that put
more spots and pores on the star, except in run ${\cal M}({\cal A}=1)$ where
brightness and magnetic flux on a hemisphere are positively correlated.

This single
snapshot of rotational modulation clearly is only one of many
instantiations of the field distribution, but gives an idea of the
magnitude of the modulations that is sufficient for the present
purpose.  The amplitude of rotational modulation for the simulation of
the Sun-like star (${\cal M}({\cal A}=1)$) at 4\,000\,\AA\ is $\sim 0.06$\%.
This single example of rotational modulation lies only somewhat below
the value of $\approx 0.1-0.2$\%\ characteristic of the active Sun
\citep{2016JSWSC...6A..18F,2016JSWSC...6A..33L}.

On the one hand, this single snapshot model is at least roughly of
comparable magnitude to observations. On the other hand, the
rotational modulation of more-active stars also comes out low if only
the emergence frequency, but not the size spectrum, of active regions
is changed.  For ${\cal M}({\cal A}=30)$, the rotational modulation at
visible wavelengths is $\sim 0.2$\%. This is considerably weaker than
what is observed in stars of correspondingly high activity. One example
is the rotational modulation observed for Kepler-17
(Table~\ref{tab:transits}) which is about 1--2\%. Other examples of
stars with comparable rotation periods can be found in
\citet{2013ApJ...771..127N} for a sample of stars that exhibit
superflaring in the {\em Kepler} data: typical amplitudes of
rotational brightness modulation for such stars fall in the range from
1\%\ to 5\%, with most clustering around 2\%.

There are two ways by which the model's rotational modulation can be
increased: raise the frequency of emergence of active regions, or
allow for larger active regions to emerge. The frequency spectrum for
emerging bipoles used in the present simulations is based on the work
by \citet{1993SoPh..148...85H}, and is thus characteristic of sunspot
Cycle~21. Without guidance on, for example, preferred longitudes for
flux emergence, statistically, one expects an increase in rotational
modulation that scales as the square root of the number of regions
\citep[see also][]{2018ApJ...853..122R}. Based on this, rather than
modifying this frequency or introducing longitudinal patterns in an
ad hoc fashion, the second possibility is explored. The largest active
regions that emerged in the 846-day period of observations analyzed by
\citet{1993SoPh..148...85H} had a total absolute magnetic flux per
polarity of roughly $4\times 10^{22}$\,Mx, with only a single occurrence of
such a large region during the period of their analysis. This led
\citet{schrijver2000} and \citet{schryver+title2001} to set
$\Phi_{\rm max}=1.5\times 10^{22}$\,Mx in their model runs to stay just
below the most uncertain part of the active-region size spectrum with
infrequent realizations on the Sun.

A later study by \citet{zhang+etal2010}, spanning a longer period,
showed regions with fluxes up to about $3\times 10^{23}$\,Mx per
polarity. When the flux-emergence spectrum is extended to this flux for
a star of otherwise solar properties (in run ${\cal M}({\cal A}=1)^+$), the rotational
modulation at visible wavelengths reaches $\sim$0.3--0.4\%. This certainly
lies in the range of solar rotational modulation, but now the
added flux leads to surface-averaged flux densities that are too high
(Table~\ref{tab:modulations}). Without experimenting further in the
present context, I hypothesize that there may be a decline in
active-region frequencies for region with fluxes approaching
$3\times 10^{23}$\,Mx, dropping below the power law seen at lower
fluxes. This would be in line with comments by \citet{zhang+etal2010},
and with the occurrence of infrequent but large regions discussed by
\citet{2012JGRA..117.8103S}, while occasionally resulting in larger
rotational modulation and also regions large enough for infrequent powerful
solar flares \citep{2013A&A...549A..66A}.

\begin{figure}
  \includegraphics[clip=true, width=9cm]{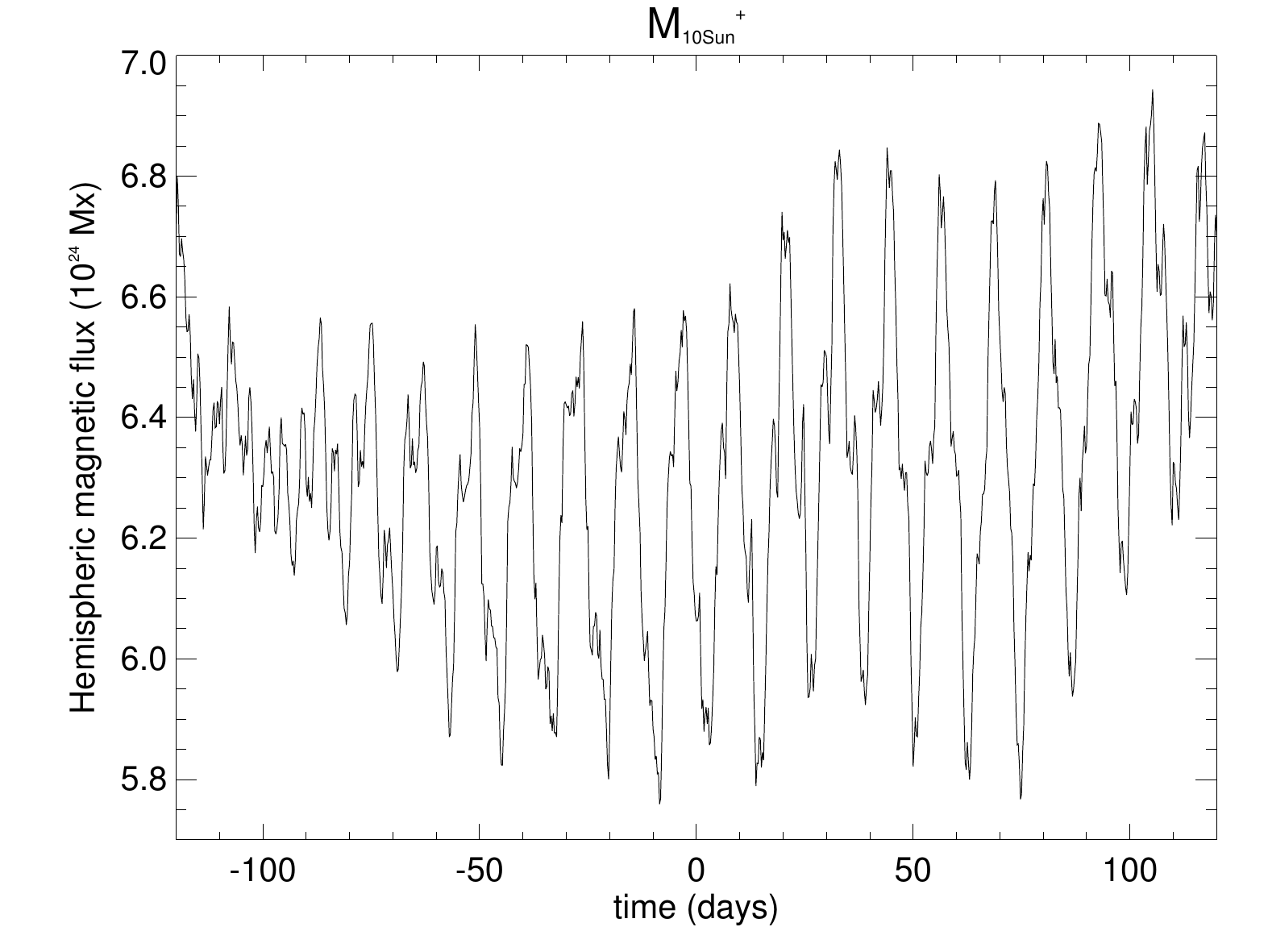}
  \caption{\label{fig:magmod} Total magnetic flux on the observer-facing side
    of the star modeled in run ${\cal M}({\cal A}=10)^+$, assuming a synodic
    rotation period of 12\,days at the Carrington reference latitude of
    $16^\circ$. The interval shown extends over 20 rotation periods,
    with the time $t=0$ set to the reference time for which
    Figures~\ref{fig:example79}c and~\ref{fig:rotmod83} are shown, at
    the full temporal resolution of the model of 6\,hr/step.}
\end{figure}
The result of allowing larger regions is also noticeable for the
more-active stars: the rotational modulation found for
${\cal M}({\cal A}=10)^+$ and ${\cal M}({\cal A}=30)^+$ with
$\Phi_{\rm max}=3\times 10^{23}$\,Mx,
is significantly larger than for
${\cal M}({\cal A}=30)$, reaching roughly 1--3\%\ in the visible. Not
only are these models in the range of the observed rotational modulation
of Kepler-17 and other similar {\em Kepler} targets, but they also
inject bipoles of sufficient magnitude to power the occasional
superflares observed on such stars
\citep[{\em e.g.},][]{2012JGRA..117.8103S,2013A&A...549A..66A,
  2013ApJ...771..127N}.

Model ${\cal M}({\cal A}=30)^{+}$ has an
average magnetic flux density of $\langle |fB| \rangle\sim 500$\,G
(Table~\ref{tab:modulations}).  \citet{2015PASJ...67...33N} show a trend
of increasing $\langle |fB| \rangle$ versus the amplitude of the
rotational brightness variation (their Figure~8b). They base this
scaling on solar observations of the Ca~II infrared triplet compared
to a magnetogram, and measure the stellar Ca~II infrared triplet to
establish the mean stellar level of activity.  If only the total flux
in facular elements (with fluxes below $3\,10^{20}$\,Mx) is taken to
contribute in the Ca~II triplet, then for model values of
$\langle |fB| \rangle_{\rm fac}\approx 180$\,G the scaling found by
\citet{2015PASJ...67...33N} would map to brightness variation
amplitudes by rotational modulation of $\approx 1.5$\%, which compares
well with what is found here (Table~\ref{tab:modulations}).

The rotational modulation is the result of the inhomogeneous
distribution of magnetic flux across the stellar surface, with the
signal presenting a mean over entire hemispheres. The result is a
relatively smooth curve as a function of the phase angle, as shown in
Figures~\ref{fig:rotmod79}--\ref{fig:rotmod80}. The hemispheric
smoothing can result in a simple modulation often at the apparent
period of rotation (Figures~\ref{fig:rotmod83}
and~\ref{fig:rotmod80}), but for the less-active stars rotation
periods can be masked. The appearance of observed light curves of
active stars may suggest the existence of only a few large spot
groups, but the images in
Figures~\ref{fig:example79}a--\ref{fig:example79}d suggest that the real
situation may be far more complex.

\begin{figure}
 \centerline{\includegraphics[width=8.2cm]{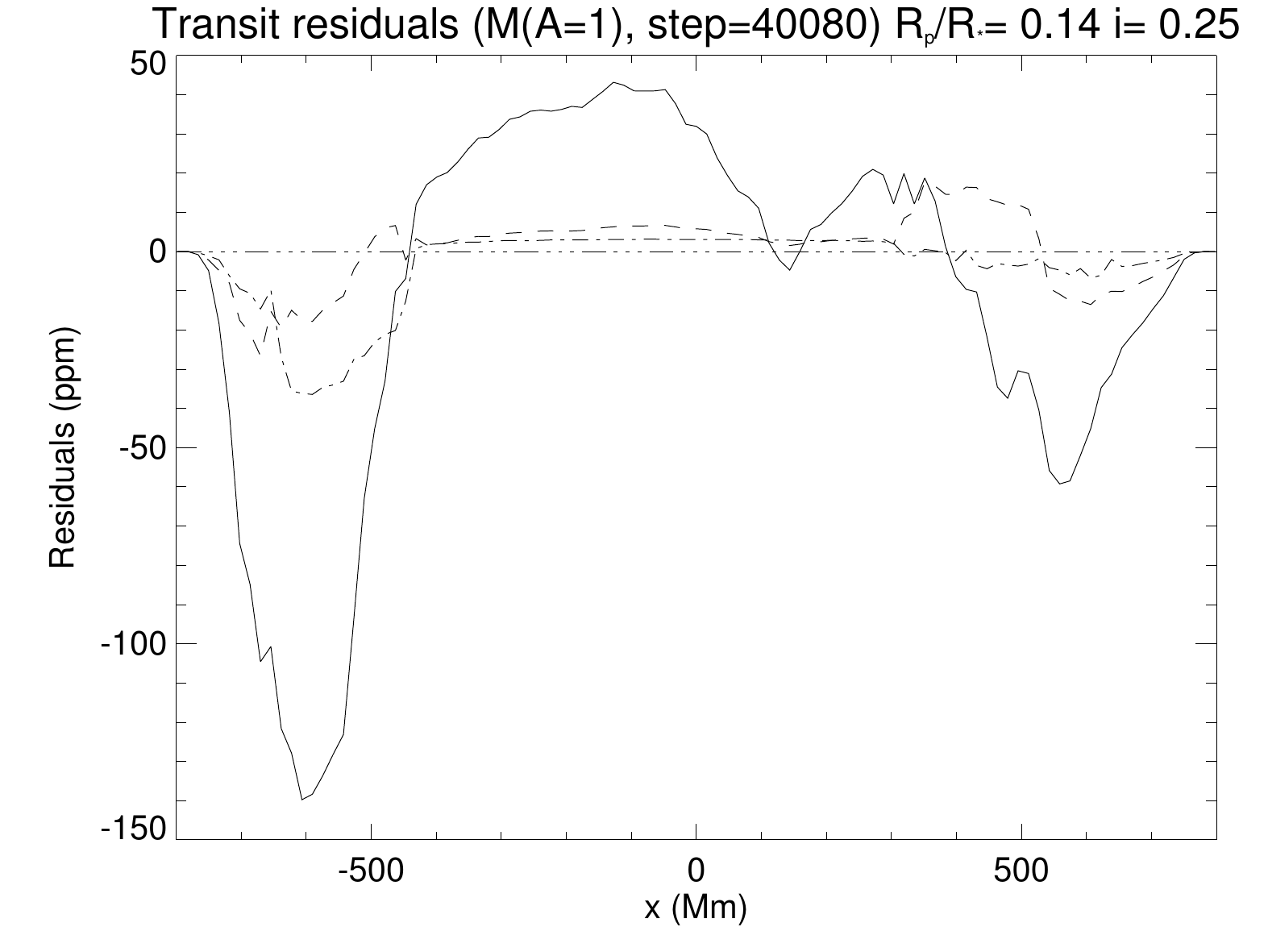}}
 \caption{\label{fig:residuals79} Transit residuals normalized to the
   out-of-transit signal for 3780\,\AA\ (solid), 6010\,\AA\ (dashed),
   and 15975\,\AA\ (dash-dotted) for an instance in
   ${\cal M}({\cal A}=1)$ (corresponding to a phase angle of $0^\circ$
   in Fig.~\ref{fig:rotmod79}. The time steps used for the transit correspond to
   $\approx 16$\,Mm in the plane of the sky.}

  \centerline{\includegraphics[width=8.2cm]{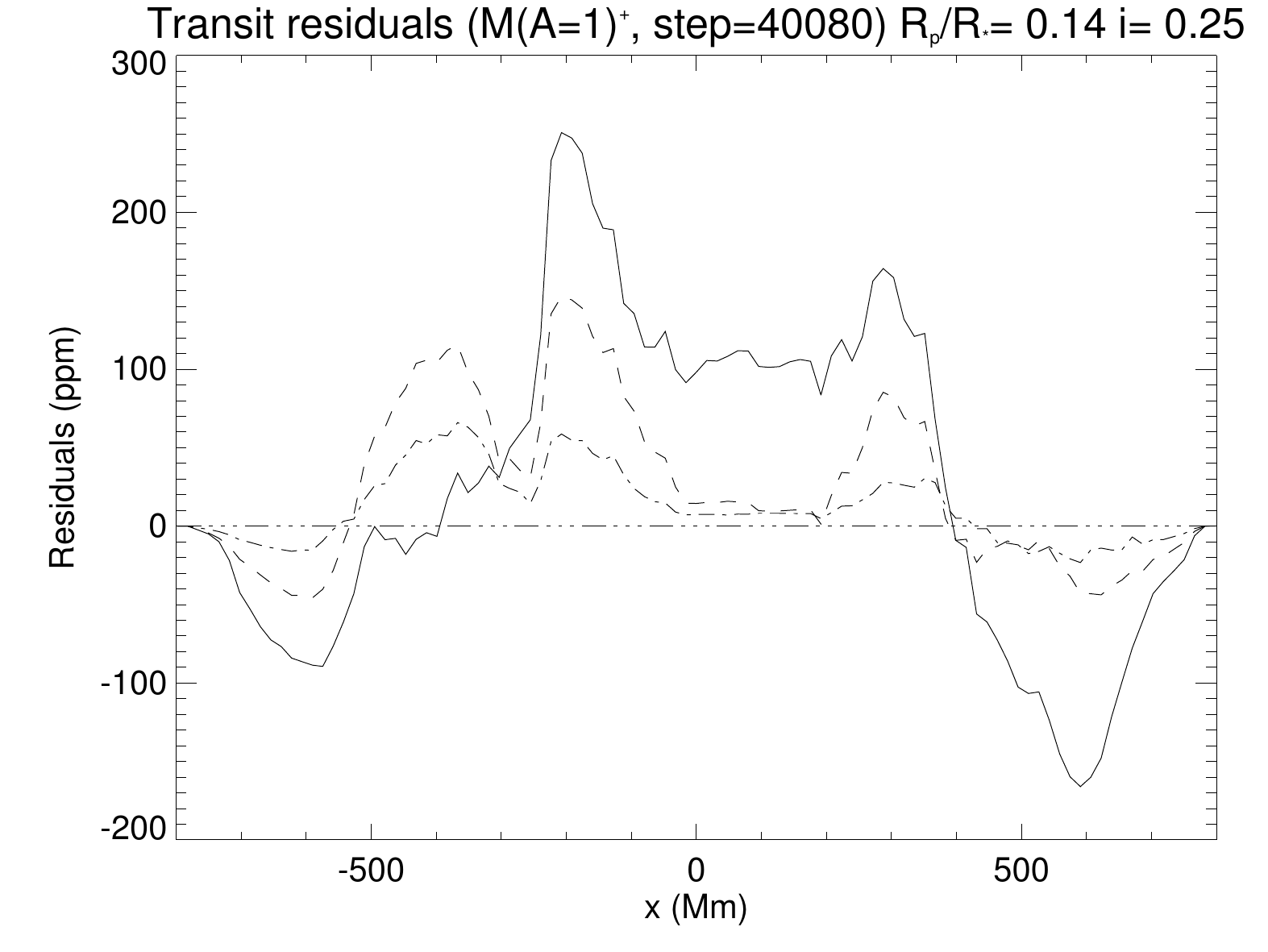}}
  \caption{\label{fig:residuals82} As Fig.~\ref{fig:residuals79} for
    ${\cal M}({\cal A}=1)^+$.}
\end{figure}
\begin{figure}
  \centerline{\includegraphics[width=8.2cm]{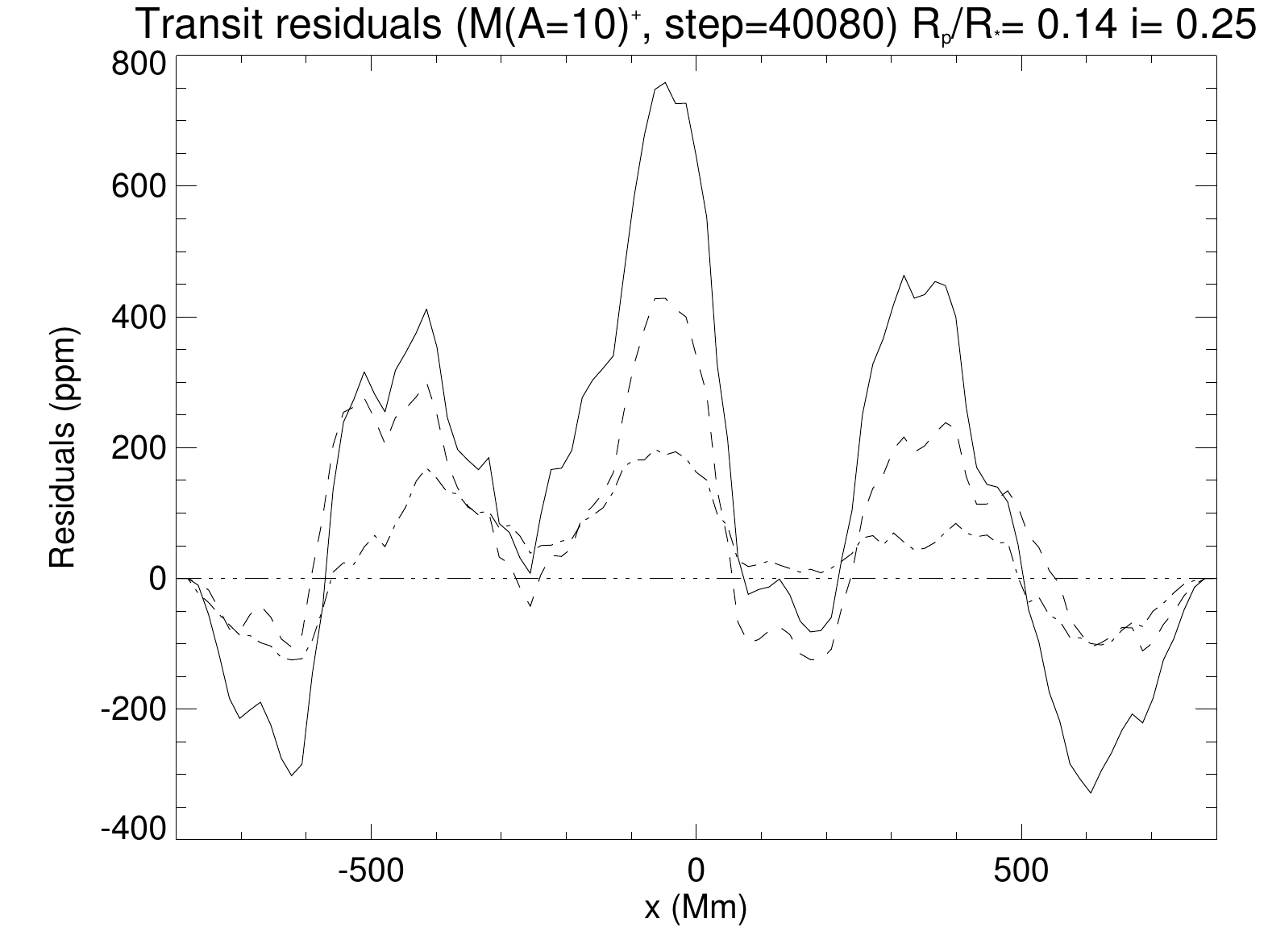}}
  \caption{\label{fig:residuals83} As Fig.~\ref{fig:residuals79} for
    ${\cal M}({\cal A}=10)^+$. Compare to
      Figure~\ref{fig:multipersp83} where a series of transits are
      shown for 3870\,\AA\ for different rotation angles of the
      simulated star.}

    \centerline{\includegraphics[width=8.2cm]{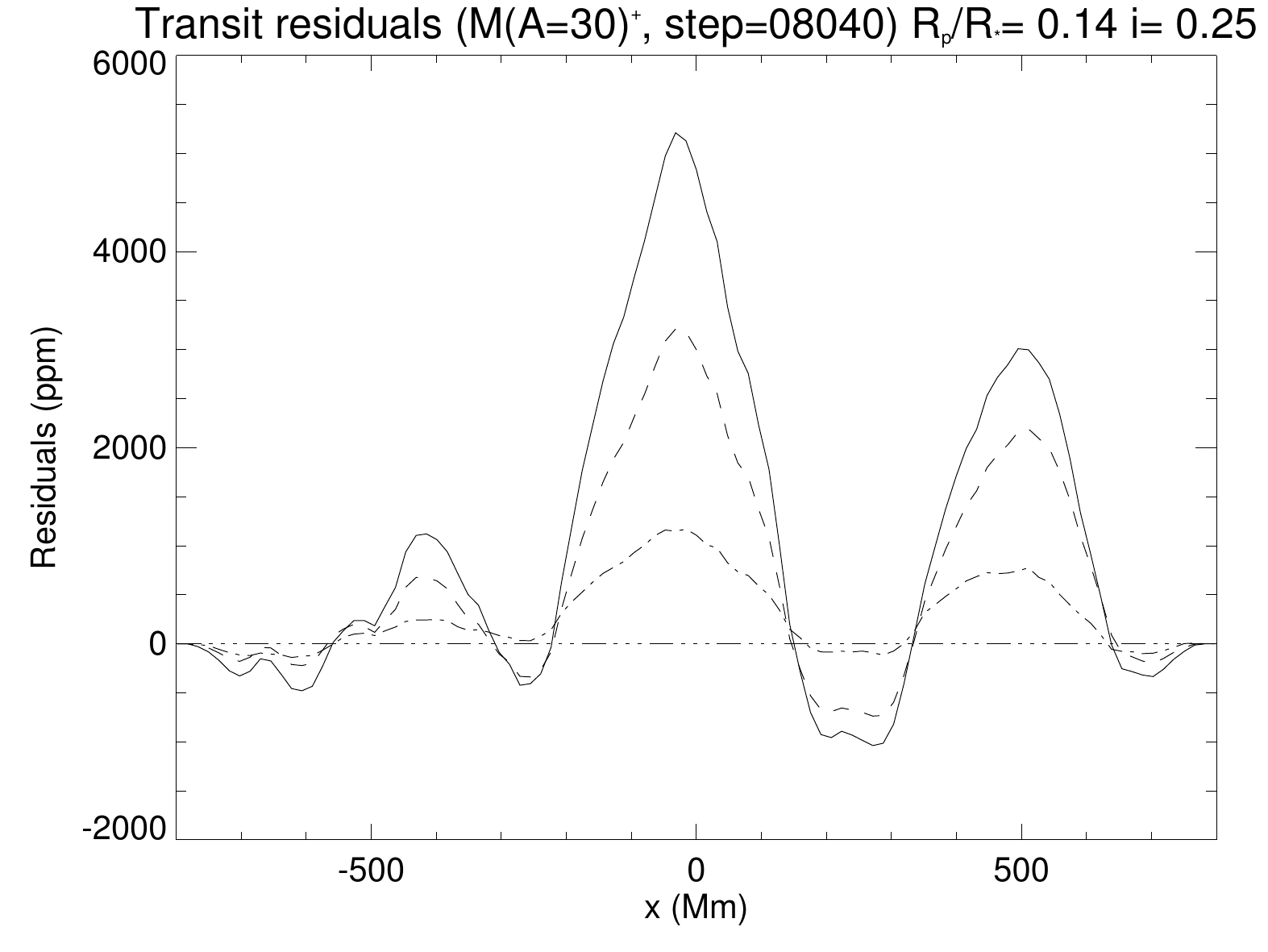}}
    \caption{\label{fig:residuals80} As Fig.~\ref{fig:residuals79} for
      ${\cal M}({\cal A}=30)^+$. }

\end{figure}
Figure~\ref{fig:magmod} illustrates the rotational modulation over
time by looking at the total amount of magnetic flux on the 'observer-facing'
hemisphere of the 'rotating star': here, the magnetic evolution is
shown for run ${\cal M}({\cal A}=10)^+$ over a total of 20 simulated rotation
periods set to 12\,days each, at the full model resolution of its 6\,hr
steps, centered on the time shown in Figure~\ref{fig:example79}c. The
figure shows a clear long-term evolution in the total amount of flux,
but for most of the period there is a pronounced rotational modulation
suggestive of a dominant active hemisphere where none is prescribed in
the model.

\section{Transit signals}\label{sec:transits}
\subsection{Aspects of model transit light curves}
The average radius of the selected exoplanets in Table~\ref{tab:transits} is
$0.14 R_\ast$, which is also characteristic of WASP-36b and
Kepler-17b, which orbit the most Sun-like stars of the longest
rotation periods in the sample listed.  In the transit simulations
discussed here, I therefore adopt a planetary radius of
$0.14 R_\odot$. Smaller virtual exoplanets can, of course, also be
used, but that is an application that can await both advanced
magnetoconvective models of stellar photospheres and very large
aperture telescopes with increased S/N properties. One
exception is made for comparison to a transit with a planetary radius of
$0.02 R_\odot$ discussed below.

\begin{figure} 
 \centerline{\includegraphics[width=8.5cm]{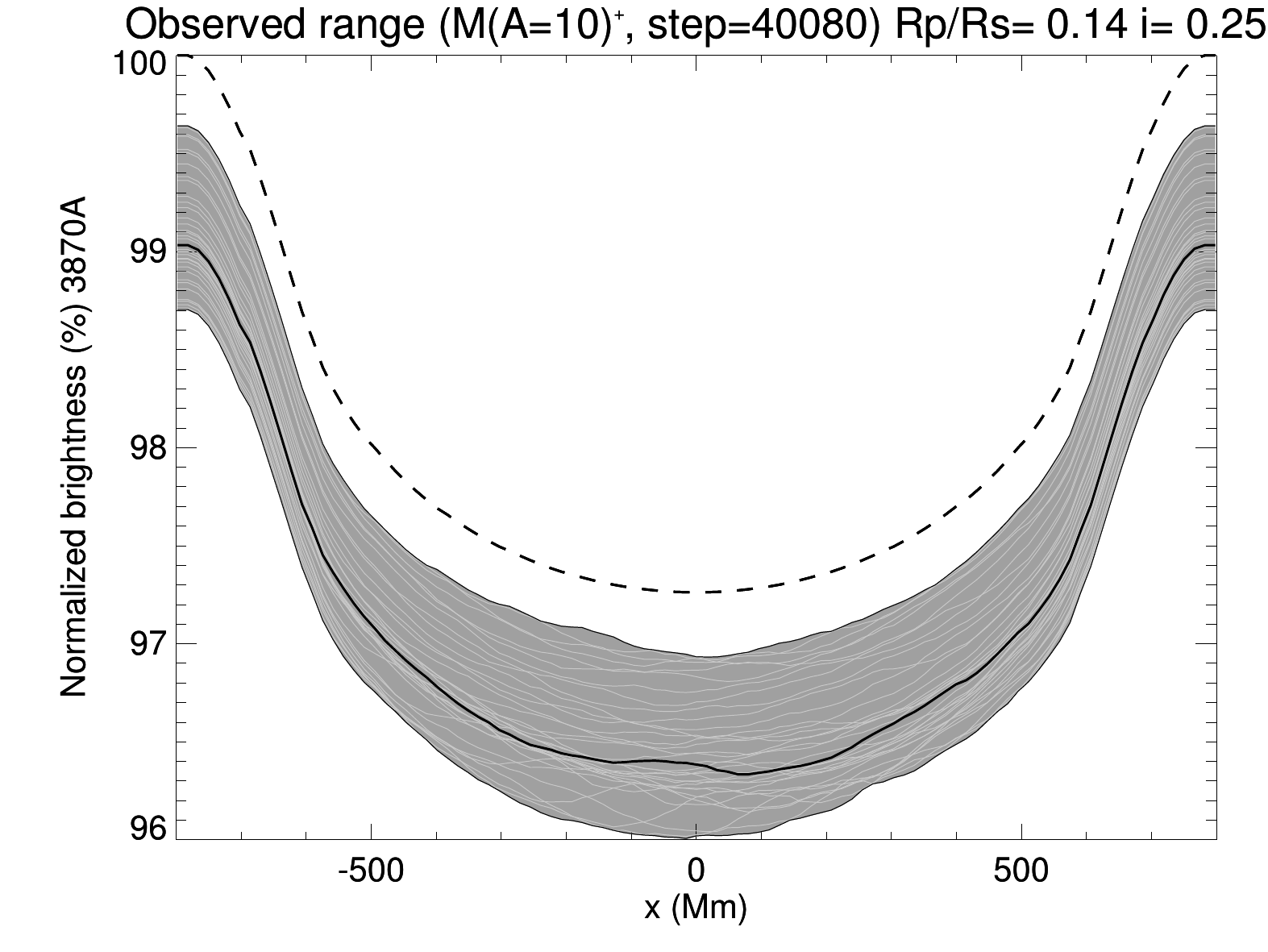}}

  \centerline{\includegraphics[width=8.5cm]{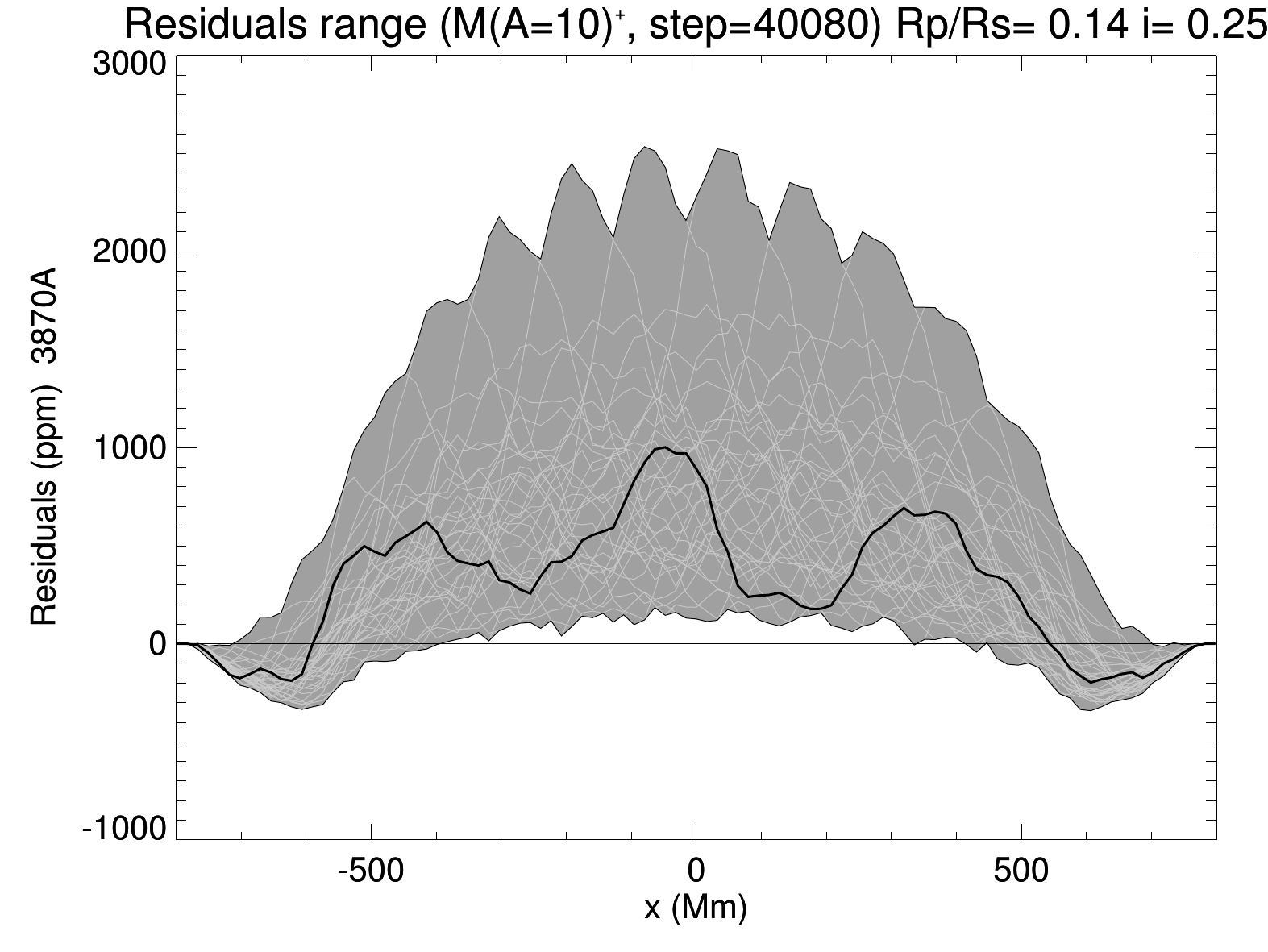}}
  \caption{\label{fig:multipersp83} (top) Transit signal and (bottom)
    transit residuals as for Figure~\ref{fig:residuals83} for run
    ${\cal M}({\cal A}=10)^+$ for 3870\,\AA\ and
    $R_{\rm p}/R_\ast=0.14$, shown by the thick black curve. Transit
    signals are computed for a series of different perspectives,
    rotating the star each time by $10^\circ$. In the top panel, all
    transit signals are normalized to the intensity of the inactive
    star. In the bottom panel, all residuals are shifted to a zero
    value out of transit. The shaded band shows the range of signals;
    note that the peaks in the upper envelope in the bottom panel are
    a result of the stepping by $10^\circ$. The light grey curves
    depict the transit residuals for each of these viewing angles. The
    dashed black line in the top diagram shows the transit signal for
    an inactive photosphere.}
\end{figure}
\begin{figure}
 \centerline{\includegraphics[width=8.5cm]{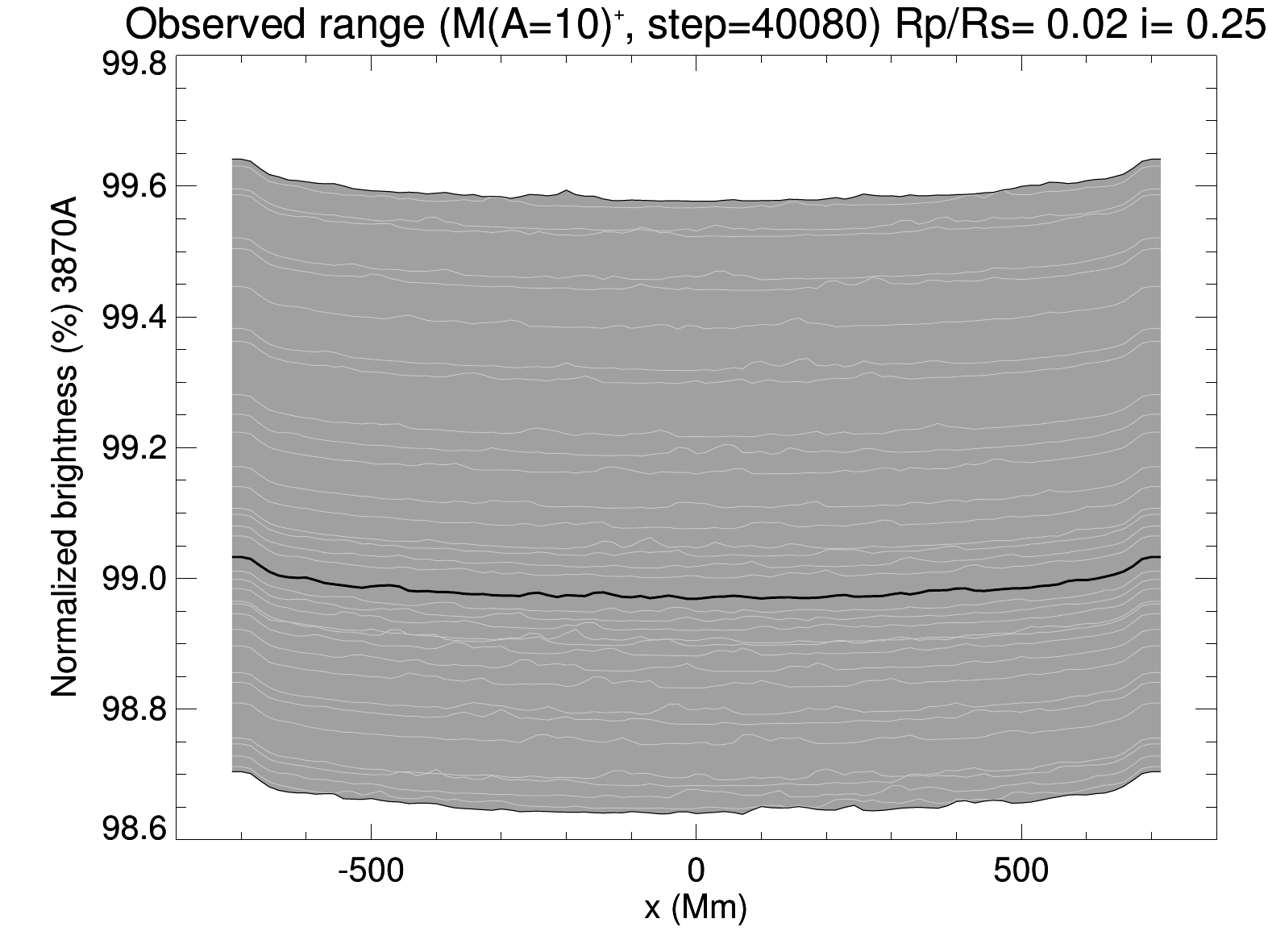}}

  \centerline{\includegraphics[width=8.5cm]{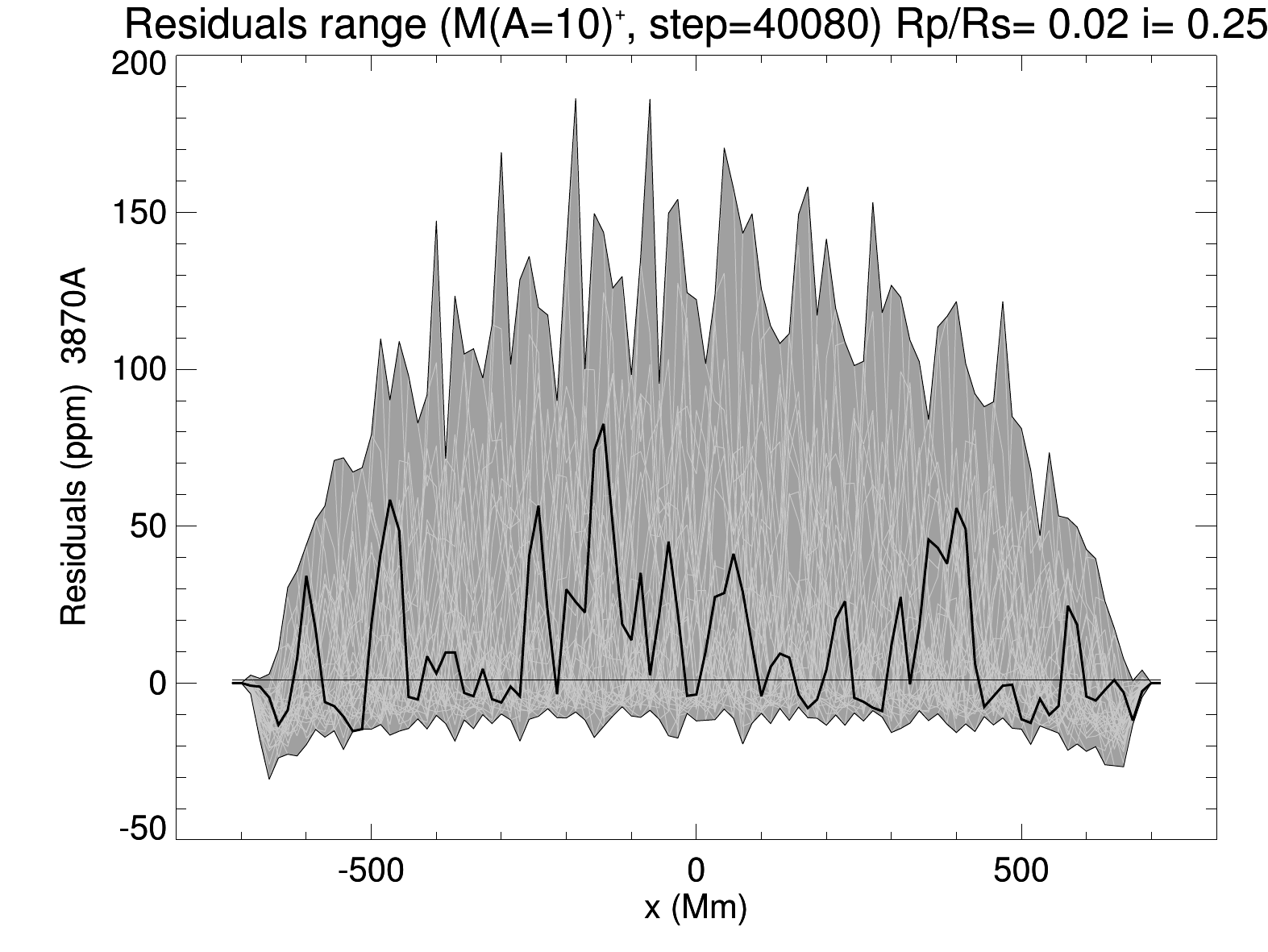}}
  \caption{\label{fig:multipersp83b} (top) Transit signal and (bottom)
    transit residuals as for Figure~\ref{fig:multipersp83}, but for a
    transiting planet with $R_{\rm p}/R_\ast=0.02$ (or $\sim
    2R_\earth$). In this case, the rotational modulation far exceeds
    the transit modulation, causing the apparently wide band in the
    top panel (note the difference in vertical scale with
    Fig.~\ref{fig:multipersp83}). }
\end{figure}
Figures~\ref{fig:example79}a--\ref{fig:example79}d show examples of
transits paths across the model stars with a relative impact parameter
for the transit of $d_{\rm t}=0.25R_\ast$ (compare with the estimated
values for the selected stellar sample in Table~\ref{tab:transits}),
for which Figures~\ref{fig:residuals79}--\ref{fig:residuals80} show
the transit residuals, calibrated relative to the inactive star and to
zero out of transit:
\begin{equation}\label{eq:transitresiduals}
{\cal R}(t,\lambda)\equiv
{(I_\ast(t,\lambda)-I_{\rm q}(t,\lambda)) \over I_{\rm q}(t,\lambda)} -
{(I_\ast(\infty,\lambda)-I_{\rm q}(\infty,\lambda)) \over I_{\rm q}(\infty,\lambda)},
\end{equation}
where $I_\ast(t,\lambda)$ is the modeled transit intensity at time $t$
(running in the figures from first contact with the image field of
view to fourth contact) and wavelength $\lambda$, and
$I_\ast(\infty,\lambda)$ is the brightness of the active star out of
transit; $I_{\rm q}$ are the same for the quiet, featureless
photosphere.  The curves are fairly smooth for two reasons: the
random-walk diffusion of the magnetic flux leads to relatively smooth
gradients in the surface field, with further smoothing imposed by the
exoplanet diameter. But there are fine-scale variations that, if they
are detectable in real observations, could be used to study
structural details of stellar magnetic fields, provided that the
effective telescope aperture is sufficiently large to keep exposures
short while obtaining a high S/N ratio.

Figures~\ref{fig:residuals79}--\ref{fig:residuals80} show only a
single instance of an arbitrarily chosen step in the flux-transport
simulations and a single viewing angle, while comparing the residuals
for the three wavelengths used here. Figure~\ref{fig:multipersp83}
provides an impression of the full range of transit curves (top panel)
and transit residuals (bottom panel) for run ${\cal M}({\cal A}=10)^+$
(compatible with a Sun-like star at $P_{\rm rot}\approx 12$\,days) for
only 3870\,\AA\ computed by rotating the star underneath a series of
transits.  The full range of transit residuals for this time step
clearly shows the effect of limb brightening by faculae in the dips
early and late in the transits. Moreover, the full range of residuals
is more than double the amplitude of the single realization selected
for Figure~\ref{fig:residuals83}. \added{For comparison,
  Figure~\ref{fig:multipersp83b} shows the simulated transit signals
  and residuals for a much smaller planet ($R_{\rm p}/R_\ast=0.02$);
  note that for such a small planet umbral-penumbral differences
  --~here ignored~-- are in principle accessible, although this would
  require S/N levels that are currently unaccessible.}

All transit residuals reveal signatures of the facular brightening
toward stellar limbs, recognized as dips near the beginnings and ends
of the transit residuals \citep[to be compared, for example, to a
similarly shaped transit light curve for Kepler-71 shown in Figure~4
of][]{2019MNRAS.484..618Z}. This is in effect a wavelength- and
activity-dependent distortion of the limb-darkening curve (which
should be anticipated to be asymmetric because the stellar activity
along the transit path is expected to be generally asymmetric).

Next, I consider the magnitude of the residuals: The peak-to-trough
ranges in the transit residuals, $\Delta(\lambda)$ (summarized in
Table~\ref{tab:modulations}) confirm the very weak transit signals
expected for a star of solar activity, ${\cal M}({\cal A}=1)$, even when
allowing for large active regions as in ${\cal M}({\cal A}=1)^+$. For the
stars listed in Table~\ref{tab:transits} with rotation periods in the
range of 8--12\,days, transit-residual amplitudes reach values of
up to 0.15--0.4\%. These values are much larger than those found for
${\cal M}({\cal A}=30)$, but are compatible with the model results for
${\cal M}({\cal A}=10)^+$ and ${\cal M}({\cal A}=30)^+$, {\em i.e.}, for
models in which much larger active regions are included in the
emergence spectrum.

Figures~\ref{fig:example79}c and~\ref{fig:example79}d reveal the
consequences of raising $\Phi_{\rm max}$ to 
$3\times 10^{23}$\,Mx: the injection of very large regions leads to a
larger length scale in the flux patterns (compared to
solar patterns as in Figure~\ref{fig:example79}a), and these large
areas of high magnetic flux densities are conducive to forming or
maintaining concentrations with high magnetic fluxes, many of which
appear as spots and pores. At least that is what happens in the
models based on the properties derived from solar observations.

\subsection{Estimating the planetary radius}
The effective area of the
occulting disk relative to that of the background star, based on the
relative occultation depth
$d(t,\lambda)=1-I_\ast(t,\lambda)/I_\ast(\infty,\lambda) $ at any time
$t$ during the transit can be expressed as follows:
\begin{equation}\label{eq:planetarea}
{R_{\rm p}^2(\lambda) \over R_\ast^2(\lambda)} = d(t,\lambda)
\left ( {I_\ast(\infty,\lambda) \over I_{\rm o}({\bf r}(t),\lambda)} \right ),
\end{equation}
where $I_{\rm o}({\bf r},\lambda)$ is the mean intensity of the patch
occulted by the transiting exoplanet
and $I_\ast(\infty,\lambda)$ is the overall mean stellar intensity out
of transit. Below, any wavelength dependence of $R_\ast$ is
ignored.

\begin{figure}
\centerline{\includegraphics[width=8.5cm]{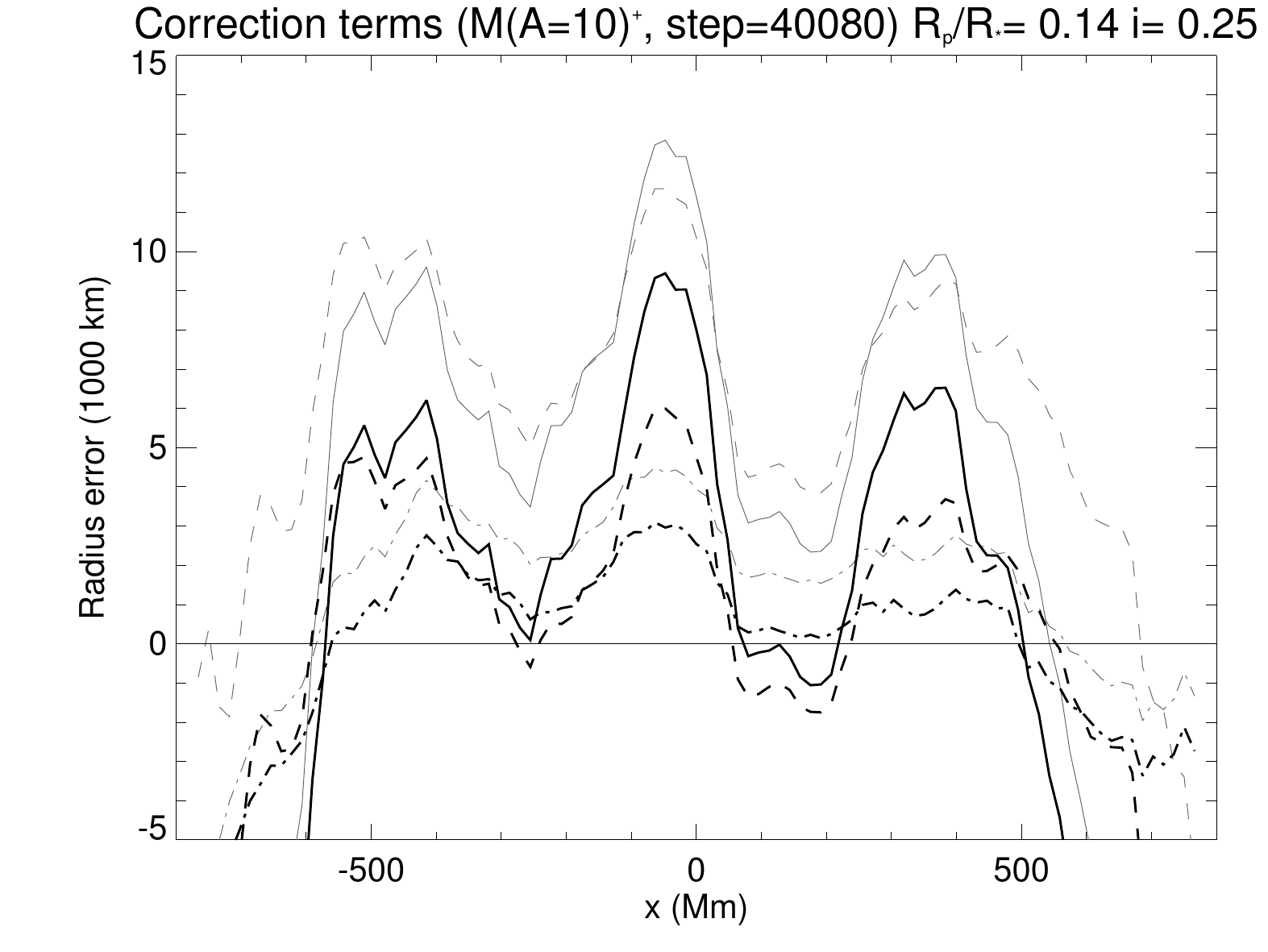}}
\caption{\label{fig:planetradiuserror} Activity-related terms for estimated
  planetary radius for a transiting planet with
  $R_{\rm p}=1.4R_{\rm Jup}$ based on the effects of stellar surface
  activity as captured in Eq.~(\ref{eq:planetradius}). The curves
  shown (as a function of the distance $x$ from central meridian
  measured against the plane of the sky) match the transit residuals as shown in
  Figure~\ref{fig:residuals79}, i.e., for model
  ${\cal M}({\cal A}=10)^+$; solid for 3870\,\AA, dashed for
  6010\,\AA, and dash-dotted for 1.6\,$\mu$m. Thin gray lines show
  the magnitude of the final term in parentheses in
  Eq.~(\ref{eq:planetradius}), while the thick black lines also
  include the overall brightness term with $\delta_\ast(\lambda)$ as shown
  in Table~\ref{tab:modulations}.}
\end{figure}
Determining the deceptively simple-looking ratio
$I_\ast(\infty,\lambda) / I_{\rm o}({\bf r}(t),\lambda)$ is a
challenge: we do not know what is masked by the exoplanet, and what is
important in the ratio is the difference between the surface
structures on the entire observer-facing side of the star and that
occulted by the planet, including limb darkening, which is relatively
poorly known compared to that to which we have some observational
access, namely along the transit path, and there is contaminated by
surface activity.

Having observations at multiple wavelengths will help in constraining
spot and plage contributions to the stellar brightness over the
spectrum, but not without more study. For example, the observed
intensities $I_\ast(\lambda)$ are related to the intrinsic
intensity of a non-active star, $I_{\rm q}(\lambda)$, the spot contrast and
filling factor, $c_{\rm s}(\lambda)$ and $f_{\rm s}$, and the facular
brightness contrast across the surface,
${\cal C}_{\rm f}({\bf r},\lambda)$, integrated over the observer-facing
hemisphere, here with total area normalized to unity:
\begin{equation}\label{eq:components1}
  I_\ast(\lambda)=I_{\rm q}(\lambda) \left(  1 + f_{\rm s} c_{\rm s}(\lambda)
    + \int {\cal C}_{\rm f} ({\bf r},\lambda) {\rm d}S \right ).
\end{equation}
If the facular contrasts would have been essentially
multiplicatively scaling functions across wavelength, i.e., if
${\cal C}_{\rm f} ({\bf r},\lambda) \approx {C_{\rm f}}({\bf
  r}){c_{\rm f}}(\lambda)$ for a contrast $C_{\rm f}$ at some
reference wavelength associated with faculae only, then
\begin{equation}\label{eq:components2}
  I_\ast(\lambda)\approx I_{\rm q}(\lambda) \left(  1 + f_{\rm s} c_{\rm
  s}(\lambda) + c_{\rm f}(\lambda) \int C_{\rm f}({\bf r}) {\rm d}S \right ),
\end{equation}
with the integral, independent of wavelength, a constant to be
determined. Then the problem of determining the unknowns would have
been a straightforward inversion problem. But the fact that the
limb-brightening curves for faculae modeled by
\citet{2017A&A...605A..45N} do not transform into one another by
simple multiplicative factors implies that in order to fully
disentangle the spot and facular components, the spatial distribution
of these components across the stellar disk needs to be reasonably
well known. For some purposes and some wavelengths, the approximation
in Eq.~(\ref{eq:components2}) may suffice, but determining this is
beyond the scope of this paper.

An estimate of the relative impact of the effect of the activity in
the transit chord on the radius of the exoplanet and any surrounding
atmosphere can be obtained by writing
\replaced{$I_{\rm o}({\bf r}(t),\lambda)=
I_{\rm q}({\bf r}(t),\lambda)+\delta I_{\rm o}({\bf r}(t),\lambda)$,
where $I_{\rm q}({\bf r}(t),\lambda)$
is the quiet, limb-darkened photosphere behind the occulting planet
and
$\delta I_{\rm o}({\bf r}(t),\lambda)$ is the (spot plus
faculae)}{$I_{\rm o}({\bf r}(t),\lambda)=
I_{\rm q}({\bf r}(t),\lambda)(1+\delta_{\rm o}({\bf r}(t),\lambda))$,
where $I_{\rm q}({\bf r}(t),\lambda)$
is the quiet, limb-darkened photosphere behind the occulting planet
and
$\delta_{\rm o}({\bf r}(t),\lambda)$ is the (spot plus faculae) relative}
difference in intensity of the occulted patch due to stellar activity.
Similarly, we can write
$I_\ast(\infty,\lambda) = I_{\rm q}(\infty,\lambda)
(1+\delta_\ast(\lambda))$.  For relatively small perturbations,
so if $\delta_{\rm o}({\bf r}(t),\lambda) << 1$ and $\delta_\ast(\lambda)<< 1$, then
Eq.~(\ref{eq:planetarea}) transforms into
\begin{eqnarray}
R_{\rm p}(\lambda) &\approx& R_\ast(\lambda)  \, d^{1\over 2}(t,\lambda)
\, \left ( I_{\rm q}(\infty,\lambda) \over I_{\rm q}({\bf r}(t),\lambda)
  \right )^{1\over 2} \times \nonumber \\
  \label{eq:planetradius}
           &&  \left ( 1+ {1\over 2} \delta_\ast(\lambda)- {1\over 2}\delta_{\rm o}({\bf
  r}(t),\lambda) \right )  .
\end{eqnarray}
The final part of this equation essentially quantifies the ``transit light
source effect'' related to stellar magnetic activity
\citep[{\em e.g.},][]{2018ApJ...853..122R,2019AJ....157...96R}.

Ignoring measurement uncertainties and assuming the quiet-star
photospheric brightness, including limb darkening, is known, the final
expression between parentheses in Eq.~(\ref{eq:planetradius})
quantifies the effect of stellar activity on the estimated planetary
radius. Figure~\ref{fig:planetradiuserror} shows an example of the
overall impact on estimated values of $R_{\rm p}(\lambda)$ for a
transiting exoplanet with radius $1.4R_{\rm Jup}$ for
${\cal M}({\cal A}=10)^+$ by the last term in the final parenthetical
expression of Eq.~(\ref{eq:planetradius}): in the near-IR, the impact is of
order 2000--4000\,km, while differences across wavelengh from the blue
to the near-IR amount to at least 1000\,km even when times of minimal
spot coverage are selected during the transit (thin gray lines in
Figure~\ref{fig:planetradiuserror}).

If the
wavelength-dependent impact of faculae and spots on overall
brightness were known, i.e., if we knew $\delta_\ast(\lambda)$ --~which we do in
this case for the models~-- then this reduces the impact on the
estimated planet radius in this example (thick black lines in
Figure~\ref{fig:planetradiuserror}), \replaced{, still leaving effects
  equivalent to some}{depending on whether the transit covers
  sufficiently inactive photospheric regions. If, however, one were to
  use the amplitude of the rotational modulation along with the
  assumption that the asymmetry between hemispheres is entirely
  attributable to a single dominant spot group, and were to ignore
  facular contributions, then the estimated values for
  $\delta_\ast(\lambda)$ would be substantially smaller than the true
  values, leading to wavelength-dependent errors equivalent to
  several} 1000\,km. For the other models in
Table~\ref{tab:modulations} these correction terms impact radius
estimates from some 500\,km in stars with solar-like activity
(${\cal A}=1$) to of order 5000\,km at the high-activity end
(${\cal A}=30$).

\section{Discussion}\label{sec:discussion}
\begin{figure}
  \includegraphics[clip=true, width=9.0cm]{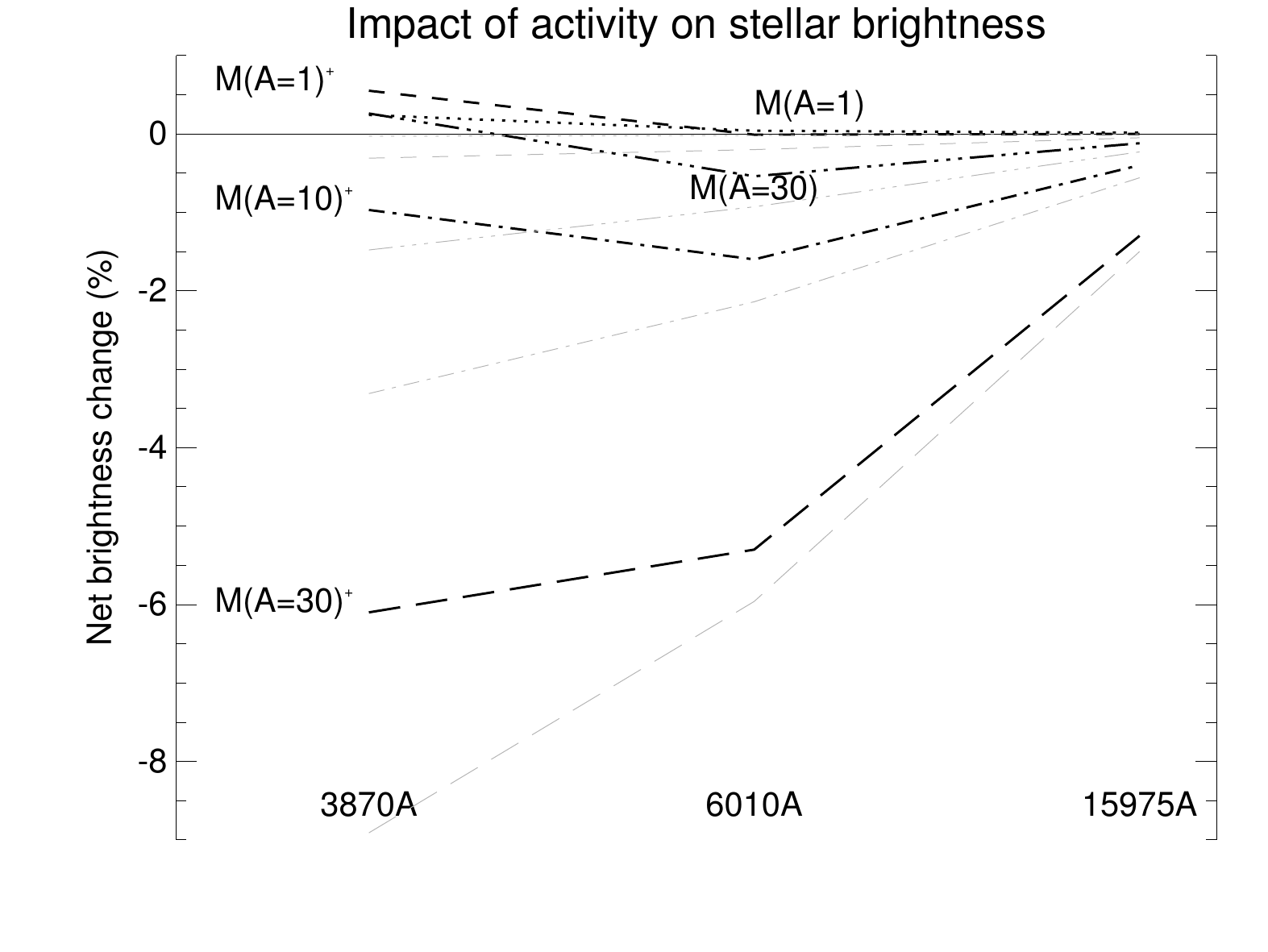}
  \caption{\label{fig:spectralsummary} Visual summary of spectral impacts (from
    Table~\ref{tab:modulations}). Plotted are the values of
    $\delta_\ast$, the ratio of the brightness of the modeled
    active star relative to that of the reference quiet
    photosphere. \added{The thick black lines show the effects of all
      surface activity, while the thin gray lines (using the same line
    patterns) include the effects of dark spots and pores alone.}}
\end{figure}

\begin{figure}
\includegraphics[clip=true, width=9.0cm]{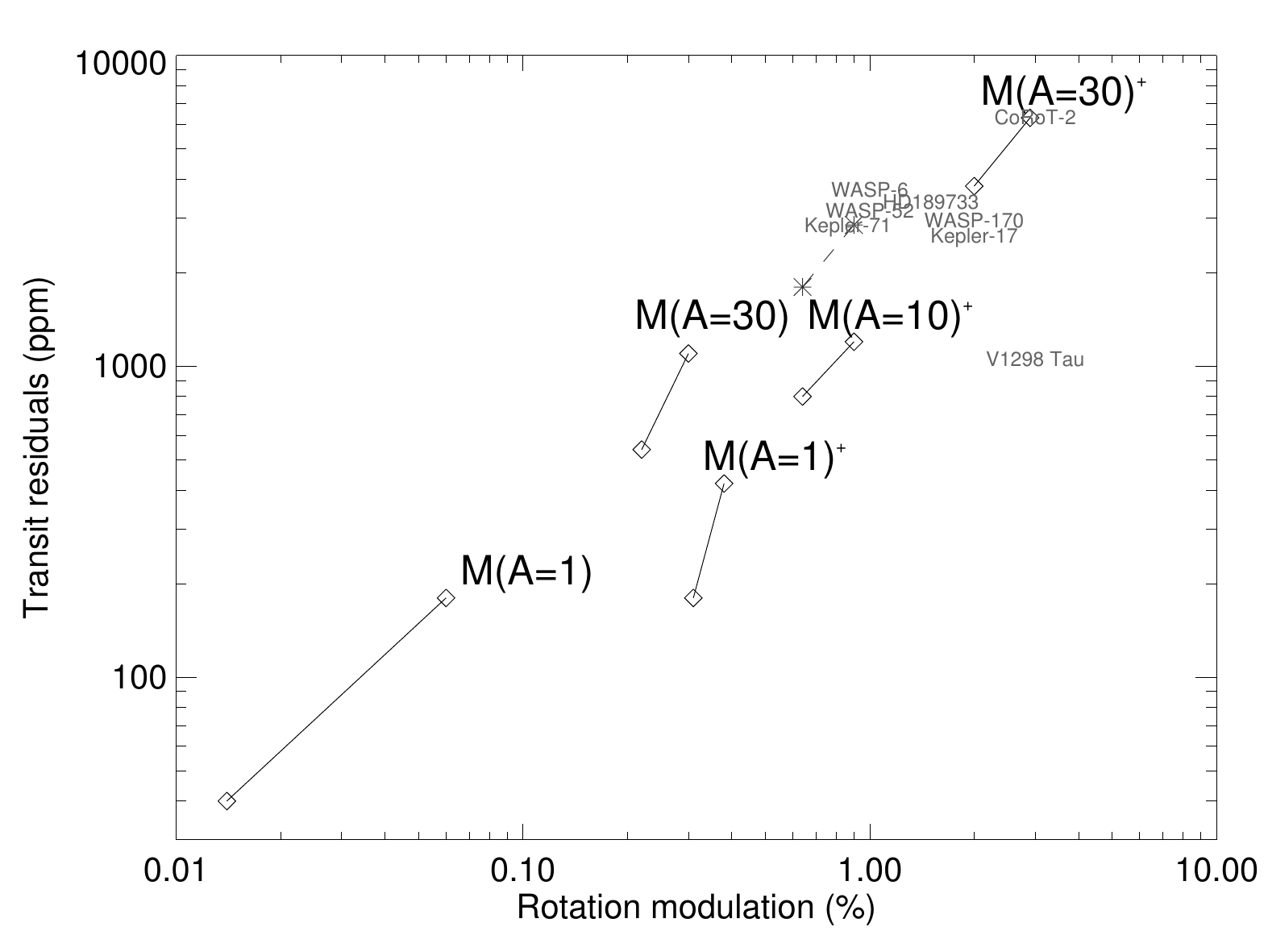}
  \caption{\label{fig:summary} Visual summary of peak-to-trough transit residuals
    vs.\ peak-to-trough rotational modulation (from
    Table~\ref{tab:modulations}). The simulation runs are shown as line
    segments, with the lower points for 6010\,\AA\ and the higher ones
    for 3870\,\AA; for run ${\cal M}({\cal A}=10)^+$ the maximum
    transit residuals expected from Figure~\ref{fig:residuals83} are
    added as asterisks connected by a dashed line segment. The stars
    for which both rotational modulation and transit residuals are
    shown in Table~\ref{tab:transits} are shown by name, roughly
    centered on the values in the table, moved slightly to reduce
    overlap where needed.}
\end{figure}

Visual summaries of the results are provided in
Figures~\ref{fig:spectralsummary}
and~\ref{fig:summary}. Figure~\ref{fig:spectralsummary} shows how the
three different wavelengths respond to surface magnetic features as a
function of modeled activity: generally, spot darkening exceeds
facular brightnening, except for stars of roughly solar activity
level, 
where facular brightening overcompensates for spot dimming,
particularly at the shortest wavelengths.

Figure~\ref{fig:summary} shows the general trend of increasing
rotational modulation with transit residuals as well as of increased
signals when larger active regions are included (both as
anticipated). On the one hand, the signals for the selected stellar
sample appear to lie somewhat above the values anticipated from the
models for corresponding levels of activity ${\cal A}\sim 10$ and
rotation periods $P_{\rm rot}\sim 12$\,days if allowing for large active
regions. On the other hand, stellar data do follow the general trend
defined by the models, \added{even, for example, in the total average
  spot coverage of $\approx 7$\%\ for Kepler-17 derived from
  rotational modeling \citep{2019AnA...626A..38L} and $\approx 5$\%\
  for WASP-52 based on a series of spectroscopic observations
  \citep{2019arXiv191105179B} compared to the values $f_{\rm d}$ in
  Table~\ref{tab:modulations}}. There are at least four possible
interpretations: (1) the selected spot temperature was set too high or
the selected average magnetic field strength to characterize
umbral-plus-penembral areas was set too low, (2) there is a selection
bias in the stellar sample (and possibly in the transit residuals
selected for display in the literature), such that the stars in the
sample were studied for their transit signals precisely because these
were strong, while on the other hand, the simulated models may have
much stronger residuals when viewed from different perspectives than
the randomly selected example that was analyzed (as is the case for
Figures~\ref{fig:multipersp83} and~\ref{fig:multipersp83b}), (3) even
larger active regions may need to be accommodated for, or (4) the
overall level of stellar activity as characterized by ${\cal A}$ from
a rough calibration to the coronal brightness (see
Sect.~\ref{sec:transport}) needs to be raised somewhat, or some mix of
the above. Possibly even the tidal effects between star and planet in
these compact systems with heavy planets is important
\citep{2000ApJ...533L.151C}. All these aspects deserve further
analysis in future studies. For now, we can also look at this diagram
as quite encouraging because of the fair agreement of models and observations
in view of the many unknowns going into the modeling.

Are the algorithmic elements of the model applicable to stars
substantially more active than the Sun and to very large active
regions?  Lacking the necessary observational spatial resolution as
well as guidance from 
magnetoconvective models, the present model results can only be tested
against the transit and rotation-modulation observations to attempt to
falsify the assumptions made.

The behavior of starspots presents one particular issue of interest. The
code does not allow for very large spots to remain intact for very long, but rather
will have such spots fragment, \replaced{to a distribution with a wide peak around
$10^{21}$\,Mx, and thus}{which, combined with collisional mergers
working the other way,} leads to the creation of clusters of dark
features. Small clusters (or individual spots) would then lead to
light-curve signatures that would be interpreted as compact and
dark. In contrast, more extended clusters of spots in the model would
lead to an observational signature that would be interpreted as a more extended and
less dark feature. Interestingly, such tendencies were reported
for the well-observed stars CoRoT-2 and Kepler-17 from exoplanet
transit modeling, as discussed in
Sect.~\ref{sec:sample}.

Rotation-modulation analyses and Doppler-imaging studies may suggest
large spotted areas, but the limited angular resolution of these
methods does not provide unambiguous information about any finer-scale
structuring within such spotted areas. The transit light-curve
analyses of CoRoT-2 and Kepler-17 discussed in Sect.~\ref{sec:sample}
is certainly suggestive of the existence of smaller starspots than
inferred from rotation modulation or Doppler imaging, and of larger
areas having smaller filling factors, causing larger spotted areas to
appear brighter.

Evidence for a population of smaller and rather uniformly distributed
population of starspots in cool stars was also inferred from
observations of members of the Pleiades cluster, a group of
$\sim 125$\,Myr-old and thus quite active cool stars:
\citet{2018ApJ...868..143G} point out that the characteristic
photometric rotational modulation is typically substantially smaller
than expected for a starspot coverage fraction deduced from TiO
absorption bands in the case of stars covered by only few large spots, but
that these observations can be reconciled if there are many smaller
starspots that are more uniformly distributed about the stars in longitude.
Similar evidence, albeit for an analysis based on mostly M-type stars,
that very active stars may be covered by multitudes of relatively
small spots rather than by a small number of larger ones also comes
from their rather weak rotational modulation, as discussed by
\citet{2013MNRAS.431.1883J}.

A question for which the answer remains simply to be determined is
whether the field dispersing from large active regions in a
photosphere much more magnetically active than the Sun's carries a
population of pores and small spots: on the Sun, the network formed by
decaying active regions is simply too weak to carry magnetic
concentrations of the required magnitude. Consequently, we do not know
from observations whether or not network concentrations on a much more
active star would form or maintain dark pores and small spots, or
whether sufficiently strong regions would function as nest sites for
newly emerging regions. Magnetoconvective experiments, however, do
support the possibility of spontaneous formation, other than upon flux
emergence, of at least pores in settings with a sufficiently high mean
magnetic flux density \citep[{\em e.g.},][]{2010ApJ...719..307K}.

Another issue that needs attention in both modeling and observations
is that of extremely large bipoles. If one takes the characteristic
rate of flux emergence of active regions (see
Sect.~\ref{sec:transport}), then the emergence of very large regions
should take many weeks to several months. This makes them appear as
persistent longitudes of activity in rotational modulation
curves. This is not inconsistent with observed stellar signals, but
one certainly needs to question how to better constrain the properties
of extremely large regions that are rarely or never observed on the Sun.

\citet{2019MNRAS.484..618Z} model transit residuals of Kepler-71
allowing for dark spot regions as well as bright facular regions. They
conclude that the facular patches have a brightness contrast of
1.1$\times$--1.25$\times$ the quiet photosphere. Such values would be
expected for areas with characteristic magnetic flux densities of
100~G or more. Extended such facular areas do occur in model
${\cal M}({\cal A}=30)^{+}$, but pores and spots within them limit the
limb brightening to less than what appears to exist on Kepler-71. The
interpretation of the relatively bright limb facular regions on the
late G-type Kepler-71 is beyond
the scope of this study.

\section{Conclusions}\label{sec:conclusions}
In summary, regarding the results from the present flux-transport
model in comparison with observations:

The flux-dispersal model based on solar empirical properties appears
consistent with stellar observations of rotational modulation and
transit residuals for Jupiter-class exoplanets, particularly when the
infrequent formation of large active regions with fluxes of a few
times $10^{23}$\,Mx (or somewhat more) is allowed for.

An extrapolation of the power-law spectrum of solar active-region
sizes to regions with corresponding fluxes of $3\times 10^{23}$\,Mx yields
results consistent with, although possibly somewhat below, observed
rotational modulation and transit residuals. Such infrequent large
regions would contain sufficient energy to power superflares that are
detected with low occurrence frequencies on moderately to very active
cool stars.

The surface flux-transport model is quantitatively consistent with a
shift from facula-dominated to spot-dominated brightness changes with
activity, as observed for cool stars.

For moderately to very active Sun-like stars, the model predicts large
area filling factors for bright faculae and dark pores and spots, such
that transits of stars substantially more active than the Sun will
rarely, if ever, cross over sizable patches of truly quiet-star
photosphere. Consequently, limb-darkening curves derived from observed
transits of active stars are not directly comparable to
convective-atmosphere models that do not incorporate magnetoconvective
effects associated with regions of high magnetic flux density. For
studies that involve passbands or spectral lines with significant
chromospheric contributions, one should expect nearly ubiquitous
spectral contamination of transit spectra by stellar features.

Facular contributions result in a wavelength-dependent change in the
observed limb darkening that elevates the spectrum increasingly
toward shorter wavelengths relative to a nonmagnetic spectrum, thus
mimicking a bluer or hotter photosphere.  Limb-darkening curves
derived by using minima of a series of transits on active stars will
be contaminated by faculae and spots, and should be used with caution
if assumed characteristic for the average quiet stellar
photosphere. Using the minima in residuals of a series of transits as
the reference from which residuals are computed causes facular
contributions to be underestimated and the inferred limb-darkening
curves to be in error.

The model results raise the possibility that large starspots inferred
from spot modeling are in fact clusters of smaller spots. This is
consistent with the size-brightness correlation seen in analyses of
several dozen transits of the active stars CoRoT-2 and Kepler-17.

The compatibility of the model results with observations is
encouraging. On the one hand, it suggests that the solar paradigm may
hold at least for early G-type stars of moderately high activity, and
that this can thus be used to learn more about stellar magnetic
activity and transit spectroscopy of exoplanetary atmospheres
alike. On the other hand, the multitude of parameters going into the
model (many of which are discussed in Sect.~\ref{sec:transport})
require testing by large-scale magnetoconvective modeling and
empirical verification, while also quiet-star limb-darkening curves
need to be established and validated. In order to expand this study to
cooler stars, of which many more are available
than are truly Sun-like stars, more information
needs to be obtained on how to implement a surface-flux transport
model at different spectral types and rotation rates
(including the many unknowns listed at the beginning of
Sect.~\ref{sec:sample}).

This work reinforces the importance of studying stellar surface
activity not only for its own sake, but also for the application of
exoplanet transit spectroscopy: differences in inferred
wavelength-dependent radii of a few hundred to a few thousand kilometers,
increasing with activity, should be anticipated as a result of stellar
surface features for stars of moderately high activity and
Jupiter-class exoplanets.

\acknowledgments This work was made possible by the support of the
International Space Science Institute in Bern, CH. I am also grateful
for the support of the US National Solar Observatory, and I thank Adam
Kowalski there, Ben Rackham at MIT, Mark Cheung at LMSAL, Jeffrey
Linsky at JILA, and an
anonymous referee for helpful comments and for suggestions to improve
the presentation.  This research has made use of the SIMBAD database,
operated at CDS, Strasbourg, France.


\end{document}